\documentclass[fleqn,usenatbib]{mnras}

\usepackage{newtxtext,newtxmath}

\usepackage[T1]{fontenc}

\usepackage{siunitx}
\usepackage{amsmath}
\usepackage{appendix}
\usepackage{comment}
\usepackage{bold-extra}

\DeclareRobustCommand{\VAN}[3]{#2}
\let\VANthebibliography\thebibliography
\def\thebibliography{\DeclareRobustCommand{\VAN}[3]{##3}\VANthebibliography}

\usepackage{graphicx}	
\usepackage{amsmath}	
\usepackage{boldline,multirow}

\usepackage{ae,aecompl}
\usepackage{comment}

\usepackage{multirow}
\usepackage{color}
\usepackage{threeparttable, tablefootnote}
\usepackage{tabularx}
\usepackage{isotope}
\usepackage[table]{xcolor}
\usepackage{pgf}
\usepackage{collcell}
\usepackage{booktabs}
\usepackage[export]{adjustbox}
\usepackage{highlight}
\usepackage{shtabularlines}
\renewcommand{\opacity}{70}
\usepackage{array}
\usepackage{multirow,bigdelim}
\usepackage{url}
\usepackage{eso-pic}

\def\reff@jnl#1{{\rm#1\/}}
\def\aj{\reff@jnl{AJ}}                 
\def\araa{\reff@jnl{ARA\&A}}           
\def\apj{\reff@jnl{ApJ}}               
\def\apjl{\reff@jnl{ApJ}}              
\def\apjs{\reff@jnl{ApJS}}             
\def\ao{\reff@jnl{Appl.Optics}}        
\def\apss{\reff@jnl{Ap\&SS}}           
\def\AAp{\reff@jnl{A\&A}}              
\def\AApr{\reff@jnl{A\&A~Rev.}}        
\def\AAps{\reff@jnl{A\&AS}}            
\def\azh{\reff@jnl{AZh}}               
\def\baas{\reff@jnl{BAAS}}             
\def\jcap{\reff@jnl{JCAP}}             
\def\jrasc{\reff@jnl{JRASC}}           
\def\mnras{\reff@jnl{MNRAS}}           
\def\pra{\reff@jnl{Phys.Rev.A}}        
\def\prb{\reff@jnl{Phys.Rev.B}}        
\def\prc{\reff@jnl{Phys.Rev.C}}        
\def\prd{\reff@jnl{Phys.Rev.D}}        
\def\prl{\reff@jnl{Phys.Rev.Lett}}     
\def\pasp{\reff@jnl{PASP}}             
\def\pasj{\reff@jnl{PASJ}}             
\def\qjras{\reff@jnl{QJRAS}}           
\def\skytel{\reff@jnl{S\&T}}           
\def\solphys{\reff@jnl{Solar~Phys.}}   
\def\sovast{\reff@jnl{Soviet~Ast.}}    
\def\ssr{\reff@jnl{Space~Sci.Rev.}}    
\def\zap{\reff@jnl{ZAp}}               
\def\nat{\reff@jnl{Nature}}            

\def\snana{\textsc{snana}}
\def\snn{SNN}

\def\DiffImg{\textsc{diffimg}}
\def\UnityClass{AllSNIa}
\def\BinInd{b}

\newcommand*{\MinNumber}{0.3}%
\newcommand*{\MaxNumber}{2.5}%

\newcommand{\ApplyGradient}[1]{%
  \pgfmathsetmacro{\PercentColor}{100.0*(#1-\MinNumber)/(\MaxNumber-\MinNumber)}%
  \edef\x{\noexpand\cellcolor{black!\PercentColor}}\x\textcolor{black}{#1}%
}
\newcolumntype{R}{>{\collectcell\ApplyGradient}{r}<{\endcollectcell}}
\definecolor{Gray}{rgb}{0.93,0.93,0.93}
\usepackage{lineno}

\defcitealias{2014AJ....147...99M}{M14}
\defcitealias{Betoule_2014}{B14}
\defcitealias{2017ApJS..233....6H}{H17}
\defcitealias{2014Ap&SS.354...89B}{Br14}
\defcitealias{2011MNRAS.412.1441L}{L11}
\defcitealias{2014AJ....147..118R}{R14}
\defcitealias{Jones_2017_I}{J17}
\defcitealias{Hlozek_2012}{H12}
\defcitealias{Shivvers_2017}{S17}
\defcitealias{2019MNRAS.485.1171K}{K19}
\defcitealias{Vincenzi_2019}{V19}
\defcitealias{Vincenzi_2020}{V21}
\defcitealias{DES-CC}{DES-CC}

\AddToShipoutPictureBG*{%
  \AtPageUpperLeft{%
    \hspace{0.75\paperwidth}%
    \raisebox{-3.5\baselineskip}{%
      \makebox[0pt][l]{\textnormal{DES-2021-0645}}
}}}%

\AddToShipoutPictureBG*{%
  \AtPageUpperLeft{%
    \hspace{0.75\paperwidth}%
    \raisebox{-4.5\baselineskip}{%
      \makebox[0pt][l]{\textnormal{FERMILAB-PUB-21-504-AE}}
}}}%

\title[Cosmological biases from photometric classification in DES-SN]
{The Dark Energy Survey Supernova Program: Cosmological biases from supernova photometric classification}

\author[DES Collaboration]{
\parbox{\textwidth}{
\Large
M.~Vincenzi,$^{1,2,3}$
M.~Sullivan,$^{1}$
A.~M\"oller,$^{4,5}$
P.~Armstrong,$^{6}$
B.~A.~Bassett,$^{8,9,10}$
D.~Brout,$^{11,12}$
D.~Carollo,$^{13}$
A.~Carr,$^{14}$
T.~M.~Davis,$^{14}$
C.~Frohmaier,$^{2,1}$
L.~Galbany,$^{15,16}$
K.~Glazebrook,$^{4}$
O.~Graur,$^{2,17}$
L.~Kelsey,$^{2,1}$
R.~Kessler,$^{18,19}$
E.~Kovacs,$^{20}$
G.~F.~Lewis,$^{21}$
C.~Lidman,$^{22,6}$
U.~Malik,$^{6}$
R.~C.~Nichol,$^{2}$
B.~Popovic,$^{3}$
M.~Sako,$^{23}$
D.~Scolnic,$^{3}$
M.~Smith,$^{1}$
G.~Taylor,$^{6}$
B.~E.~Tucker,$^{6}$
P.~Wiseman,$^{1}$
M.~Aguena,$^{24}$
S.~Allam,$^{25}$
J.~Annis,$^{25}$
J.~Asorey,$^{7}$
D.~Bacon,$^{2}$
E.~Bertin,$^{26,27}$
D.~Brooks,$^{28}$
D.~L.~Burke,$^{29,30}$
A.~Carnero~Rosell,$^{24}$
J.~Carretero,$^{31}$
F.~J.~Castander,$^{15,16}$
M.~Costanzi,$^{32,33,34}$
L.~N.~da Costa,$^{24,35}$
M.~E.~S.~Pereira,$^{36,37}$
J.~De~Vicente,$^{7}$
S.~Desai,$^{38}$
H.~T.~Diehl,$^{25}$
P.~Doel,$^{28}$
S.~Everett,$^{39}$
I.~Ferrero,$^{40}$
B.~Flaugher,$^{25}$
P.~Fosalba,$^{15,16}$
J.~Frieman,$^{25,19}$
J.~Garc\'ia-Bellido,$^{41}$
D.~W.~Gerdes,$^{42,36}$
D.~Gruen,$^{43}$
G.~Gutierrez,$^{25}$
S.~R.~Hinton,$^{14}$
D.~L.~Hollowood,$^{39}$
K.~Honscheid,$^{44,45}$
D.~J.~James,$^{11}$
K.~Kuehn,$^{46,47}$
N.~Kuropatkin,$^{25}$
O.~Lahav,$^{28}$
T.~S.~Li,$^{48,49}$
M.~Lima,$^{50,24}$
M.~A.~G.~Maia,$^{24,35}$
J.~L.~Marshall,$^{51}$
R.~Miquel,$^{52,31}$
R.~Morgan,$^{53}$
R.~L.~C.~Ogando,$^{35}$
A.~Palmese,$^{54}$
F.~Paz-Chinch\'{o}n,$^{55,56}$
A.~Pieres,$^{24,35}$
A.~A.~Plazas~Malag\'on,$^{57}$
K.~Reil,$^{30}$
A.~Roodman,$^{29,30}$
E.~Sanchez,$^{7}$
M.~Schubnell,$^{36}$
S.~Serrano,$^{15,16}$
I.~Sevilla-Noarbe,$^{7}$
E.~Suchyta,$^{58}$
G.~Tarle,$^{36}$
C.~To,$^{59,29,30}$
T.~N.~Varga,$^{60,61}$
J.~Weller,$^{60,61}$
and R.D.~Wilkinson$^{62}$
\begin{center} (DES Collaboration) \end{center}
}
}
\date{$\star$ maria.vincenzi@duke.edu\\ Author affiliations are shown in Appendix \ref{aff}\\
Accepted XXX. Received YYY; in original form ZZZ\\}

\pubyear{2021}

\begin{document}
\label{firstpage}
\pagerange{\pageref{firstpage}--\pageref{lastpage}}
\maketitle

\maketitle
\begin{abstract}
Cosmological analyses of samples of photometrically-identified Type Ia supernovae (SNe Ia) depend on understanding the effects of \lq contamination\rq\ from core-collapse and peculiar SN Ia events. We employ a rigorous analysis on state-of-the-art simulations of photometrically identified SN Ia samples and determine cosmological biases due to such \lq non-Ia\rq\ contamination in the Dark Energy Survey (DES) 5-year SN sample. As part of the analysis, we test on our DES simulations the performance of SuperNNova, a photometric SN classifier based on recurrent neural networks. Depending on the choice of non-Ia SN models in both the simulated data sample and training sample, contamination ranges from 0.8--3.5 per cent, with the efficiency of the classification from 97.7--99.5 per cent. Using the Bayesian Estimation Applied to Multiple Species (BEAMS) framework and its extension BBC (\lq BEAMS with Bias Correction\rq), we produce a redshift-binned Hubble diagram marginalised over contamination and corrected for selection effects and we use it to constrain the dark energy equation-of-state, $w$. Assuming a flat universe with Gaussian $\Omega_M$ prior of $0.311\pm0.010$, we show that biases on $w$ are $<0.008$ when using SuperNNova and accounting for a wide range of non-Ia SN models in the simulations. Systematic uncertainties associated with contamination are estimated to be at most $\sigma_{w, \mathrm{syst}}=0.004$. This compares to an expected statistical uncertainty of $\sigma_{w,\mathrm{stat}}=0.039$ for the DES-SN sample, thus showing that contamination is not a limiting uncertainty in our analysis. 
We also measure biases due to contamination on $w_0$ and $w_a$ (assuming a flat universe), and find these to be $<$0.009 in $w_0$ and $<$0.108 in $w_a$, hence 5 to 10 times smaller than the statistical uncertainties expected from the DES-SN sample. 

\end{abstract}


\section{Introduction}

Type Ia supernovae (SNe Ia) are widely used in cosmology to directly measure the accelerating expansion rate of the universe, and to characterise the properties of the \lq dark energy\rq\ thought to cause it. Following the original detection of the accelerating cosmic expansion using SNe Ia \citep{1998AJ....116.1009R, 1999ApJ...517..565P}, two decades of  time-domain surveys have discovered and followed up thousands of cosmologically-useful SNe Ia, from the local universe to redshifts beyond $z\sim1$. As the statistical power of these samples has improved, there has been a commensurate reduction in systematic uncertainties that has broadly tracked the increase in SN Ia numbers \citep{2006A&A...447...31A, 2009ApJS..185...32K, 2011ApJ...737..102S, 2014A&A...568A..22B,2014ApJ...795...44R, scolnic2018, 2018ApJ...853..126R,DES_abbott}. However, unlocking the full constraining power of current and future samples of SNe Ia requires a new level of controlling systematic uncertainties introduced by the use of photometric SN classification. Modelling and assessing systematic biases introduced by SN classification is the main focus of this paper.

Photometric SN classification methods are needed when candidate SNe detected by a survey lack a spectroscopic confirmation of their type. In these cases, most cosmological analyses to date have been restricted to SN events with spectroscopic redshift from the likely host galaxy, and SN classification is based on the characteristics of the observed light curve. Early approaches were frequently used for individual high-redshift SN events forming part of relatively small samples \citep[e.g.,][]{1999ApJ...517..565P,2007ApJ...659...98R}, albeit often using other contextual information such as host galaxy type. More general approaches include selecting candidate SNe Ia based on their light-curve fit properties \citep{2011A&A...534A..43B} and classifying SNe based on both template fitting (e.g., pSNID, \citealp{2011ApJ...738..162S,2018PASP..130f4002S} or \citealp{2014ApJ...795..142G}) and machine-learning approaches \citep{Lochner_2016,2016JCAP...12..008M,2020MNRAS.491.4277M}.

The outputs from SN photometric classifiers require a careful interpretation, as instead of the simple binary classification associated with spectroscopic classification (i.e., SN Ia or not a SN Ia), photometric classifiers return the probability of each event being a SN Ia, $P_\mathrm{Ia}$. A framework 
is needed to marginalise over the contamination from events that are not SNe Ia. The Bayesian Estimation Applied to Multiple Species (BEAMS) method \citep[][]{2007PhRvD..75j3508K}, and its extension \lq BEAMS with Bias Corrections\rq\ \citep[BBC;][]{Kessler_2017}, are frequently used in this context, the latter also incorporating corrections due to selection effects based on high-quality survey simulations.

The development of photometric classification has been motivated by the recent and future large SN surveys like the Sloan Digital Sky Survey (SDSS) SN Survey \citep{2018PASP..130f4002S}, the Pan-STARRS Medium Deep Survey \citep{Jones_2017_I,Jones_2018_II}, the Dark Energy Survey SN program \citep{2012ApJ...753..152B, DES_spec} and the future Legacy Survey of Space and Time \citep[LSST]{2019ApJ...873..111I}. These SN imaging surveys motivated large spectroscopic follow-up programs to measure host-galaxy redshifts for the majority of discovered SNe, and use them for cosmological measurements.
The first measurement of the equation-of-state of dark energy, $w$, with a photometric SN Ia sample was performed by \citet{2013ApJ...763...88C} using data from the SDSS SN Survey. They used pSNID, together with a selection of events based on their SN Ia light-curve fit properties, which together reduced contamination in the SN Ia sample to an estimated 3.9 per cent. However, the systematic effects of this contamination on the final measurement of $w$ was not estimated. \citet{Hlozek_2012} first demonstrated the application of BEAMS on the SDSS SN sample \citep[similar to the sample used by][]{2013ApJ...763...88C}, but also lacked an assessment of systematic uncertainties in the analysis.

The cosmological analysis of the Pan-STARRS (PS1) photometric SN sample \citep{Jones_2017_I, Jones_2018_II} was the first to include an evaluation of the cosmological biases and systematic uncertainties introduced by contamination in the photometrically-classified SN Ia sample. Using several simple classification approaches that don't rely on machine learning, including pSNID, the biases on measurements of $w$ due to contamination were estimated to be small, and the associated systematic uncertainty was estimated to be $\sigma_{w,\mathrm{syst}}=0.012$. This uncertainty is significantly smaller than the total systematic uncertainty on $w$ of 0.043, illustrating that, under the assumptions of this analysis, contamination resulted in a small contribution to the total uncertainty budget.

Recently developed photometric classifiers \citep{Lochner_2016, 2020MNRAS.491.4277M} have shown a good performance on simulated samples of SNe developed for various classification challenges \citep{2010PASP..122.1415K,2019PASP..131i4501K,2020arXiv201212392H}. However, a critical issue remains: the training and validation of these classifiers are often performed on the same sample of simulated SN events. These simulated samples are generated either applying the same selection function of the test set, or assuming the training sample is biased towards brighter events due to spectroscopic selection effects. These simulations may not reflect the true diversity of the transient universe, and may require tuning in their input astrophysics to reproduce the observed characteristics of the selected SN sample \citep{Jones_2017_I,Jones_2018_II}. This procedure can potentially lead to an over-estimation of the classifier performance and thus underestimate systematic uncertainties in measured cosmological parameters. Ultimately, the development of accurate SN survey simulations for the training and validation of these photometric classifiers is at least as important as the development of the classifiers themselves.

This paper investigates biases in the measurement of cosmological parameters that are introduced in the use of photometric SN classification algorithms within the BBC framework. Our focus is on the Dark Energy Survey\footnote{\url{https://www.darkenergysurvey.org/}} (DES) SN program \citep[DES-SN;][]{DES_spec} dataset. DES-SN is a state-of-the-art sample for SN Ia cosmology analysis, with approximately 2000 likely SNe Ia in the final \lq 5-year\rq\ sample: $\sim20$ per cent of the SNe have follow-up spectroscopy of the SN itself \citep[e.g.,][]{DES_spec}, and most of the remaining events have a host galaxy spectroscopic redshift \citep[see][]{lidman2020ozdes}.

\citet[][hereafter \citetalias{Vincenzi_2020}]{Vincenzi_2020} previously presented large simulations of DES-SN that generate realistic samples of transients that accurately describe DES-SN data. The simulation includes the \lq normal\rq\ SNe Ia, improved core-collapse SN spectral templates \citep[][hereafter \citetalias{Vincenzi_2019}]{Vincenzi_2019} and peculiar SNe Ia \citep[SN1991bg-like SNe and SN2002cx-like SNe;][]{2019PASP..131i4501K}, as well as the DES survey characteristics, to make accurate predictions for the expected populations of SNe in DES-SN. These simulations demonstrated an excellent agreement between data and simulated SN properties across many parameter distributions, including Hubble residuals and Hubble residual distribution tails. Analysing these simulated samples in detail, and fitting all the detected events with the SALT2 SN Ia light-curve model \citep{Guy_2007}, \citetalias{Vincenzi_2020} predicted 6--8 per cent of the sample to be comprised of events that are not SNe Ia, after an event selection based on the light-curve properties and fitted SALT2 parameters. No photometric classification algorithm was used.

Here we generate simulations as in \citetalias{Vincenzi_2020} to assess the performance of the SuperNNova (\snn) photometric SN classifier \citep{2020MNRAS.491.4277M} when applied to DES-SN data. \snn\ is a deep learning classifier that identifies SNe Ia with high accuracy \citep[see analyses presented by][ and M{\"o}ller et al. in prep.]{2020MNRAS.491.4277M}. We exploit the BEAMS implementation in the BBC framework to assess the impact of contamination on the cosmological analysis of the DES-SN photometric sample. The strength of our analysis lies in the fact that we use realistic simulations of SNe Ia and non-Ia SN contamination, that have been shown to reproduce the general photometric properties of the DES-SN data to high accuracy \citepalias{Vincenzi_2020}. We also test the effect of a range of astrophysically-plausible core-collapse SN model variations on the final cosmological measurements.

The paper is outlined as follows. In Section~\ref{sec:des}, we review the DES-SN data set and the simulation infrastructure used in our analysis. Section~\ref{sec:cosmo_framework} details our cosmological analysis framework, including distance estimation, BEAMS, and bias corrections. Section~\ref{sec:photometric-classification} introduces the \snn\ classifier and assesses its performance on our simulated datasets, and in Section~\ref{sec:cosmo_bias} we present an analysis of the cosmological biases introduced by the photometric classification of the DES-SN sample. We conclude in Section~\ref{sec:conclusions}.

\section{DES-SN data and simulations}
\label{sec:des}

\begin{table*}
   \caption{Summary of core-collapse SN assumptions in the DES-SN simulations.}
   \label{table:sims}
 \centering
\begin{tabular}{lccccc}
\hline
Label & Template & Luminosity	&Dust & Avg number of SNe after & Percentage of\\
 & library & functions	& model & light-curve selection $^{\dagger}$ & Ia, PecIa, II, Ibc\\
\hline 
   Baseline & \citetalias{Vincenzi_2019} &  revised \citet{2011MNRAS.412.1441L}, Gaussian parameterization & N/A$^*$ & 1650 & 93.4, 1.4, 4.5, 0.8\\

LFs+Offset &\citetalias{Vincenzi_2019} &  revised \citet{2011MNRAS.412.1441L} + 0.5 mag brightening offset & N/A & 1722 & 90.5, 1.4, 6.8, 1.3\\

Dust(H98)  & dereddened \citetalias{Vincenzi_2019} &  revised \citet{2011MNRAS.412.1441L}, Gaussian parameterization & H98$^{\ddag} $ & 1687 & 93.2, 1.4, 4.2, 1.1\\

J17  & \citetalias{Jones_2017_I} &  adjusted LFs from \citet{2011MNRAS.412.1441L} & N/A & 1667 & 94.3, 1.4, 3.0, 1.4\\ 

DES-CC  & \citetalias{DES-CC} &  \citetalias{DES-CC} & N/A & 1687 & 91.6, 1.4, 0.5, 6.5\\ 
\hline
    \end{tabular}
    \begin{tablenotes}
        \item $^*$N/A: not applicable: core-collapse SN templates are not corrected for host galaxy extinction, and the simulation does not include extinction.
        \item $\dagger$ Selection criteria from Section~\ref{sec:salt2_cut}, without classification. Numbers are calculated as the mean over 50 realizations of the DES-SN survey. Each simulation include SNe Ia, peculiar SNe Ia and core-collapse SNe. Normal SNe Ia alone account for 1522 events on average. 
        \item $\ddag$ \citet{1998ApJ...502..177H}.
    \end{tablenotes}
\end{table*}

DES is an optical imaging survey designed to constrain the properties of dark energy and other cosmological parameters by combining four different astrophysical probes: weak gravitational lensing, large scale structure, galaxy clusters and SNe~Ia \citep{2019PhRvL.122q1301A}. DES ran for six years and used the Dark Energy Camera \citep[DECam;][]{2015AJ....150..150F}, mounted on the Blanco 4-m telescope at the Cerro Tololo Inter-American Observatory. For time-domain science, DES monitored ten 3-deg$^2$ fields with an average cadence of 7 days in the $griz$ filters. Eight of the ten fields were surveyed to a depth of $\sim23.5$\,mag per visit (\lq shallow fields\rq), and the remaining two to a deeper limit of $m\sim24.5$ mag per visit (\lq deep fields\rq), thus extending to $z\sim$1.2 the redshift limit to detect SNe Ia. 

\subsection{The DES photometric SN sample}
The primary goal of the DES-SN programme is to measure the light curves of a sample of SNe Ia for use in cosmological analyses. In this paper, we use the same DES photometric SN sample as described in \citetalias{Vincenzi_2020}. This sample includes $\sim$3,600 events that have an identified host galaxy and accurately measured host galaxy spectroscopic redshift, and that pass light-curve quality selection: observations in two filters with at least one epoch with a signal-to-noise ratio (SNR) $>5$, at least one observation before the estimated time of peak brightness, and one observation after ten days (rest-frame) after peak brightness.

Following \citetalias{Vincenzi_2020}, SN host information is derived from the deep coadded images of \citep{DES_deepstacks}, and SN light-curve photometry measured using the DES Difference Imaging pipeline \citep[\DiffImg,][]{2015AJ....150..172K}. The quality of the \DiffImg\ light curves is adequate for the analysis presented in this paper, but we highlight that the final DES SN light-curves with a more accurate and precise scene modelling photometry (SMP) approach \citep{2013A&A...557A..55A,DES_SMP} is in the process of being applied to all DES-SN data. We also note that approximately 200 new host galaxy spectroscopic redshifts have been processed and incorporated into the sample while this analysis was developed. However, in this work we use the \citetalias{Vincenzi_2020} sample to maintain consistency with that analysis. 

\subsubsection{Low-$z$ SN sample}
\label{sec:lowz_data}
As this paper considers the cosmological impact of our modelling choices and photometric classification methods, we include a \lq low-$z$\rq\ (i.e., $z<0.1$) external SN Ia sample to combine with our DES-SN sample. We include five publicly available low-$z$ samples from the Harvard-Smithsonian Center for Astrophysics \citep[CfA3S, CfA3K, and CfA4;][]{2009ApJ...700..331H, 2012ApJS..200...12H}, the Carnegie Supernova Project \citep[CSP-1;][]{2010AJ....139..519C} and the Foundation Supernova sample \citep[DR1][]{Foley_Foundation}. These samples include spectroscopically confirmed SNe Ia only, therefore they are not affected by contamination.

\subsection{Simulations}
\label{sec:simulations}
Our SN simulations use SN time-series spectrophotometric templates, rates, luminosity functions and empirical relationships between SNe and their host galaxies, as well as the DES survey characteristics, to simulate the transient populations detected in the five years of DES-SN. The simulations are presented in detail in \citetalias{Vincenzi_2020} and are generated using the supernova analysis software package \citep[\snana;][]{Kessler_2009} as described in \citetalias{Vincenzi_2020}. The simulation and analysis code were orchestrated by the \textsc{pippin} \citep{Hinton2020}\footnote{\url{https://github.com/Samreay/Pippin}} pipeline. 

\citetalias{Vincenzi_2020} presented nine DES-SN simulations testing different modelling choices and assumptions.
The analysis presented in this paper has been tested for the full set of simulations presented in \citetalias{Vincenzi_2020}. However, for simplicity we focus on a reduced sample of five simulations, that encapsulate a wide range of scenarios and provides the most informative results. These simulations are:
\begin{itemize}
\item \lq Baseline\rq\: a simulation built using the core-collapse SN templates of \citetalias[][]{Vincenzi_2019}, and luminosity functions presented by \citet{2011MNRAS.412.1441L} and revised as described by \citetalias{Vincenzi_2019};
\item \lq LFs+Offset\rq\: same as Baseline, but with the core-collapse SN luminosity functions brightened by 0.5\,mag;
\item \lq Dust(H98)\rq\: uses the host-galaxy dust extinction-corrected core-collapse SN templates of \citetalias{Vincenzi_2019}, using the revised \citet{2011MNRAS.412.1441L} luminosity functions and a dust distribution presented by \citet*{1998ApJ...502..177H};
\item \lq J17\rq\: uses the core-collapse SN templates of \citet[][hereafter J17]{Jones_2017_I} together with their adjusted luminosity;
\item \lq DES-CC\rq\ simulations: uses a new set of core-collapse templates of 
Hounsel et al. in prep. (hereafter, \citetalias{DES-CC}), built from a magnitude-limited sample ($i<21.5$) of spectroscopically and photometrically identified non type Ia SNe from DES-SN.
\end{itemize}
The main characteristics of each simulation are summarized in Table~\ref{table:sims}. We also consider two simulation subsets, one that includes only SNe Ia and one that includes only SNe Ia and peculiar SNe Ia (\lq Only pec Ia\rq). These subsets exclude exclude core-collapse SNe, and are used to disentangle the effects of core-collapse SN contamination from other sources of systematic biases in the analysis. 

In all DES-SN simulations, host galaxies are associated with SNe using published SN rates as a function of global galaxy properties (stellar mass and star formation rate). We use separate rates for SNe Ia, peculiar SNe Ia, stripped envelope SNe (type Ib, type Ic and type IIb SNe) and hydrogen-rich SNe \citepalias[type II and type IIn SNe; see section 4.5 in][]{Vincenzi_2020}. We also include the dependence of the SN Ia light-curve shape on host galaxy properties.

We combine the DES-SN simulations with simulations of the low-$z$ SN Ia samples introduced in Sec.~\ref{sec:lowz_data}. These samples are simulated following \citet[][section 7.2]{2019MNRAS.485.1171K} and \citet[][section 3.1]{Jones_Foundation} and simulate mocks of the CfA (CfA3S, CfA3K, CfA4), CSP-1 and the Foundation Supernova samples.

For both the DES-SN and low-$z$ simulations, we assume the SN Ia intrinsic brightness in rest-frame $B$-band to be $M_B=-19.365$ and we set the nuisance parameters applied for stretch and colour corrections, $\alpha$ and $\beta$, equal to $\alpha=0.167$, $\beta=3.1$. Moreover, we use a flat \lq $\Lambda$ cold dark matter\rq\ ($\Lambda$CDM) cosmological model as input, with a Hubble constant $H_0=70$\,km\,s$^{-1}$\,Mpc$^{-1}$ and $\Omega_\mathrm{M}=0.311$ \citep[e.g.,][]{collaboration2018planck}. We generate 50 realisations of the DES-SN survey and pair these with 50 realisations of the low-$z$ sample. Throughout, the statistical properties of the simulated samples are presented as the mean of the 50 realisations, and uncertainties are measured as the standard deviation.

\begin{table*}
\caption{Number of observed and simulated SNe following the application of various selection criteria.}
\label{table:cuts}
\begin{tabular}{cccc|ccc}
\hline
Selection criteria & \multicolumn{3}{c}{Data} & \multicolumn{3}{c}{Simulations (avg over 50 realizations $^{c}$)}\\
& DES-SN & Low-$z$ &Total & DES-SN & Low-$z$ & Total \\
 \hline

SALT2 selection & 1676 & 312 & 1995 & 1650 & 400 & 2050\\
SALT2 selection + valid bias correction$^{a}$ & 1603 & 288 & 1891 & 1588  & 380  & 1969 \\
SALT2 selection $+$ Chauvenet's criterion$^{b}$ & 1561 & 309 & 1870 & 1572 & 400 & 1972\\
SALT2 + valid bias corr $+$ Chauvenet  &1533 & 286 & 1819 & 1545 & 380 & 1926\\
SALT2 + valid bias corr $+$ Chauvenet $+$ SALT2 $c<$0.15 & 1353 & 273 & 1626 & 1336 &  361 & 1697 \\

\hline
\end{tabular}
    \begin{tablenotes}\footnotesize
        \item $^{a}$ See Section~\ref{sec:biascor} for the definition of valid bias corrections.
        \item $^{b}$ See Section~\ref{sec:outlier_rejection} for a discussion about Chauvenet's criterion and outlier rejection methods.
        \item $^{c}$ Number of SNe averaged over 50 realizations ($N_{\mathrm{SNe}}$). The typical r.m.s. measured over the 50 realizations is $\sqrt{N_{\mathrm{SNe}}}$.
    \end{tablenotes}
\end{table*}

\subsection{SN light-curve fitting and selection}
\label{sec:salt2_cut}

We fit all simulated and observed SN light-curves with the SALT2 SN Ia light-curve model \citep{Guy_2007, Guy_2010} using the trained model parameters from \citet{Betoule_2014} and a $\chi^2$-minimization program in \snana. This fit determines several rest-frame parameters under the assumption that the event is a SN Ia: the time of SN peak brightness $t_0$, a stretch-like \citep{1997ApJ...483..565P} parameter $x_1$, a colour parameter $c$ and the light-curve normalisation parameter $x_0$, as well as their uncertainties (i.e., $\sigma_{t_0}$, etc.). We select SN events in both simulations and data that are well described by this SALT2 model. This selection is based on the fit parameters, their uncertainties, and the goodness of the light-curve fit (\lq FitProb\rq \footnote{FitProb $\in$ [0,1] and is the computed probability from $\chi^2$ and number of degrees of freedom, and assuming Gaussian-distributed errors. It quantifies how well each light curve is described by the SALT2 model.}). This is the same selection as used in \citetalias{Vincenzi_2020} and in the Joint Light-Curve Analysis sample \citep[JLA;][]{Betoule_2014}. In detail, the selection requirements are:
\begin{itemize}
    \item $|x_1|<3$ and $|c|<0.3$,
    \item $\sigma_{x_1}<1$ and $\sigma_{t_0}<2$ days,
    \item $\text{FitProb}>0.001$.
\end{itemize}
The outcome of applying this selection to our data and simulations can be found in Table~\ref{table:cuts}. The result is a data sample of 1676 SNe from DES-SN and 312 low-$z$ SNe (155 SNe from the CfA and CSP samples and 157 from the Foundation sample). Averaging our 50 Baseline simulations, we have $1650$ 
SNe from DES-SN and $400$ 
at low-$z$ ($161$ 
SNe Ia from the CfA and CSP samples, and $238$ 
SNe Ia from Foundation).

We also explore a tighter selection on the SN colour $c$, removing redder SNe using a selection of $-0.3<c<0.15$. This further reduces contamination from core-collapse SNe, with a minimal and easy-to-model loss of SNe Ia (see Table~\ref{table:cuts}). This asymmetric colour selection is also motivated by the fact that several analyses have shown that redder SNe Ia exhibit larger scatter on the Hubble diagram \citep{BS20, Kelsey2020}.

\section{Cosmological Analysis Framework}
\label{sec:cosmo_framework}

\begin{figure*}
\centering
\includegraphics[width=\linewidth]{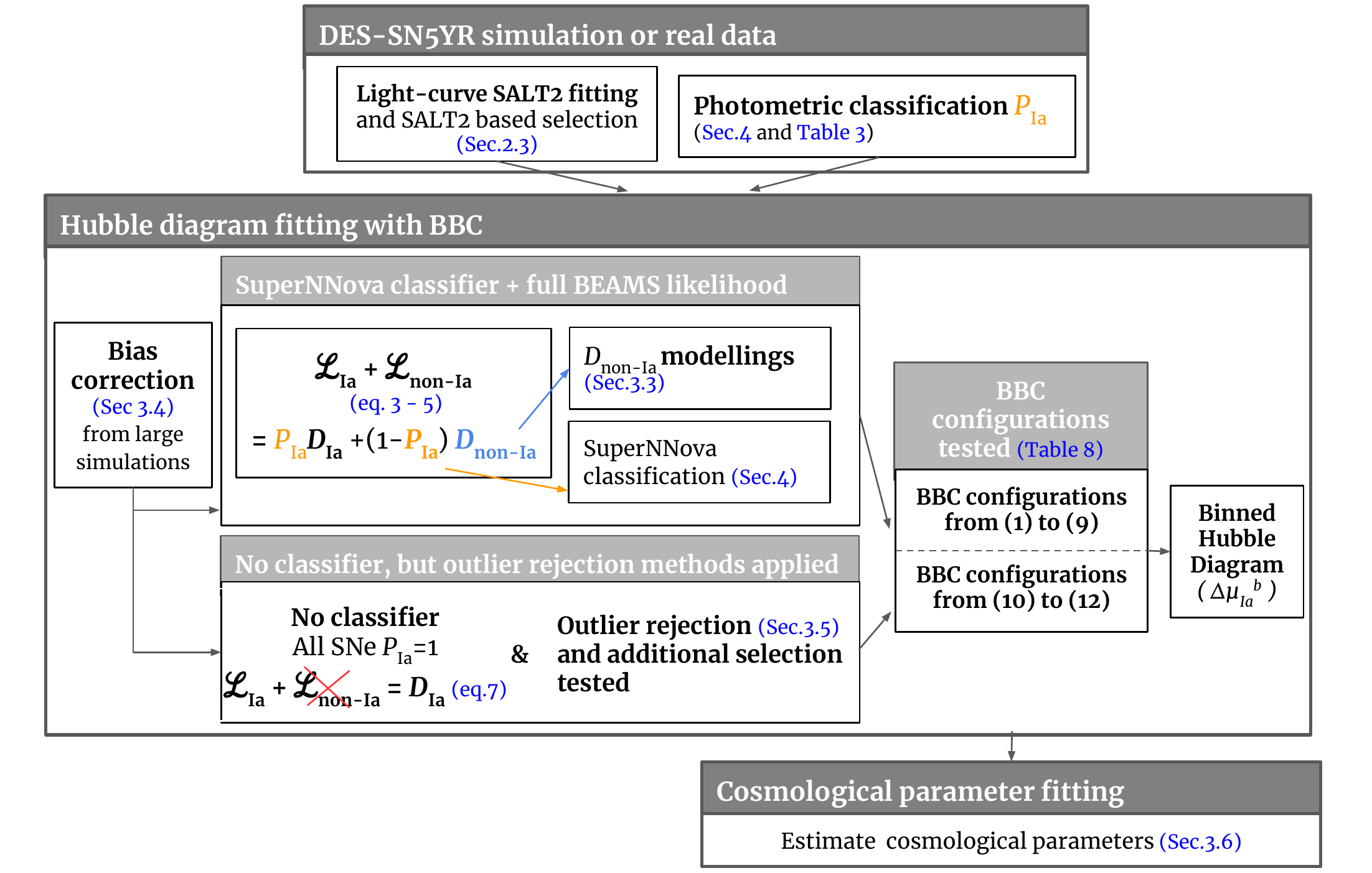}
\caption{Flow chart of the cosmological analysis framework BBC \citep[][]{Kessler_2017}, exploited in this work. BBC is specifically designed to estimate cosmological parameters from samples of photometrically identified SNe Ia. Photometric classifiers are introduced in Section~\ref{sec:photometric-classification}, while the different BBC configurations tested in this work are listed in Table~\ref{table:BBC_opts} and discussed in Section~\ref{sec:cosmo_bias}.}
    \label{fig:flow_chart}
\end{figure*}

Next, we briefly review the framework used to measure the SN Ia redshift--distance relation (\lq Hubble diagram\rq) and estimate cosmological parameters from our SN data and simulations. We begin by describing the method used to estimate distances from the SN Ia light curve parameters (Section~\ref{sec:distance-estimation}). We then present the Hubble diagram fitting method called \lq BEAMS with Bias Corrections\rq\ \citep[BBC;][]{Kessler_2017}. In the BBC method, we implement \textit{(i)} the method presented by \citet{2011ApJ...740...72M} to determine SN distances and nuisance parameters (Section~\ref{sec:distance-estimation}), \textit{(ii)} the BEAMS formalism \citep[][]{Kunz_2012} to marginalize over the contamination from non-Ia SNe (Section~\ref{sec:BEAMS}), and \textit{(iii)} simulated bias corrections to account for survey selection effects (Section~\ref{sec:biascor}). The main output of the BBC framework is a redshift-binned SN distance--redshift relation corrected for selection effects and core-collapse SN contamination, from which the cosmological parameters can be estimated (Section~\ref{sec:cosmo_estimation}). BBC also produces fitted nuisance parameters (Section~\ref{sec:salt2_cut}).
The cosmological analyses framework discussed in this section is illustrated in Fig.~\ref{fig:flow_chart}. 

\subsection{Distance estimation}
\label{sec:distance-estimation}
 
The SN Ia distance modulus, $\mu_\mathrm{obs}$, is  \citep[e.g.,][]{1998A&A...331..815T, 2006A&A...447...31A}
\begin{equation}
    \mu_\mathrm{obs} = m_B + \alpha x_1 - \beta c + \mathcal{M}_B + \Delta\mu_\mathrm{bias},
    \label{eq:tripp}
\end{equation}
where $m_B=-2.5 \log_{10}(x_0)$ and $\mathcal{M}_B$ is the absolute magnitude of a SN Ia with $x_1=0$ and $c=0$. The global nuisance parameters $\alpha$ and $\beta$ are determined following the approach presented by \citet{2011ApJ...740...72M}, i.e., fixing the cosmological parameters to some reference values (e.g., $\Omega_M=0.3$, $w=-1$) and fitting for distance modulus offsets, $\Delta\mu^{\BinInd}$, evaluated at different (log-spaced) redshift bins. A correction, $\Delta\mu_\mathrm{bias}$, is applied to each SN to correct for selection effects from the survey and analysis (see Section~\ref{sec:biascor}). 

We neglect the dependence between $\mu_\mathrm{obs}$ and host galaxy properties in our simulations and fitting \citep[e.g.,][]{2010MNRAS.406..782S}. These correlations can shift the dark energy equation-of-state $w$ by approximately one per cent \citep{DES_massstep} but ignoring them has negligible impact on studies of systematics related to contamination.

\subsection{The BEAMS likelihood}
\label{sec:BEAMS}

BEAMS is a Bayesian framework for using photometric classifications of SNe Ia, and their probabilities, in cosmology. The BEAMS likelihood requires for each SN an estimate of its probability of being a SN Ia, $P_{\text{Ia}}$. This set of probabilities are generally determined using photometric classifiers.

The BEAMS formalism is implemented in BBC, and used to fit for a binned Hubble diagram. We define the binned Hubble diagram as a set of binned distance modulii, $\mu_{\mathrm{Ia}}^{\BinInd}$, evaluated for each of the $N_{\mathrm{bins}}$ redshift bins.\footnote{We note that this binned Hubble diagram $\mu_{\mathrm{Ia}}^{\BinInd}$ is distinct from the distance modulus for individual events in equation~\ref{eq:tripp}.}
The binned distance modulii $\mu_{\mathrm{Ia}}^{\BinInd}$ are estimated by maximazing the BEAMS likelihood. This is defined as the sum of two terms, one that models the SN Ia population, $\mathcal{L}_\mathrm{Ia}$, and the other that models a population of contaminants,
\begin{equation}
    \sum_{i=1}^{N_{\mathrm{SNe}}} (\mathcal{L}^i_\text{Ia}+\mathcal{L}^i_\text{non-Ia}).
     \label{eq:tot_likelihoods}
\end{equation}
The two terms of the likelihood, $\mathcal{L}^i_\text{Ia}$ and $\mathcal{L}^i_\text{non-Ia}$, are defined as
\begin{equation}
  \begin{aligned}
    \mathcal{L}^i_\text{Ia} &=  \tilde{P}^i_{\text{Ia}} \times \text{exp} \biggl(-\frac{(\mu_{\text{obs},i} + \Delta \mu^{\BinInd} - \mu_{\text{ref}}(z_i))^2} {\sigma_{\mu,i}^{2}} \biggr)\\
    \mathcal{L}^i_\text{non-Ia} &= (1-\tilde{P}^i_{\text{Ia}}) \times D_{\text{non-Ia}}(z_i, \mu_{\text{obs},i}, \mu_{\text{ref},i}).
  \end{aligned}
  \label{eq:likelihoods}
\end{equation}
where $\mu_{\text{ref}(z_i)}$ is the distance modulus of the $i$-th SN as predicted assuming a fixed reference cosmology ($\Omega_M=0.3$, $w=-1$), and $\Delta\mu^{\BinInd}$ are the offsets quantifying by how much observations deviate from the reference cosmology in each redshift bin. By construction, the binned Hubble diagram, $\mu_{\mathrm{Ia}}^{\BinInd}$ is equal to $ \mu_{\text{ref} (\langle z \rangle _{\BinInd})} - \Delta\mu^{\BinInd}$. The distance modulus uncertainties $\sigma_{\mu,i}$ include the uncertainties propagated from the SALT2 light-curve fit ($\sigma_{m_B}$, $\sigma_{x_1}$, $\sigma_{c}$ and relative covariances), the intrinsic SN Ia scatter ($\sigma_{\text{Ia, int}}$) and peculiar velocity corrections uncertainties. The SN Ia intrinsic scatter term is determined as discussed by \citet[][section 5.5]{Kessler_2017}.

In equation~\ref{eq:likelihoods}, the terms $\tilde{P}^i_{\text{Ia}}$ and (1-$\tilde{P}^i_{\text{Ia}}$) are weighting factors applied to the two likelihoods, and represent the \lq scaled\rq\ probabilities of the $i$-th SN being a SN Ia and a core-collapse SN or peculiar SN Ia respectively. The scaled probabilities are defined as:
\begin{equation}
  \begin{aligned}
    \tilde{P}^i_{\text{Ia}} & = \frac{P^i_{\text{Ia}} }{P^i_{\text{tot}}} \text{\hspace{3mm} and \hspace{3mm} }
    \tilde{P}^i_{\text{non-Ia}} = \frac{S_{\text{non-Ia}} (1-P^i_{\text{Ia}}) }{P^i_{\text{tot}}}\\
    P^i_{\mathrm{tot}} & = (P^i_{\text{Ia}} + S_{\text{non-Ia}} (1-P^i_{\text{Ia}}))
\end{aligned}
\label{eq:renormalization_pre}
\end{equation}
where $P^i_{\text{Ia}}$ is the probability of the $i$-th SN being a SN Ia as predicted by a classifier, and $S_{\text{non-Ia}}$ is a scaling factor and an additional free parameter in the minimization of the likelihood. This additional factor enables correcting for inaccurate probabilities\footnote{Photometric classifiers often do not provide calibrated probabilities.} and it is equal to one for perfectly calibrated probabilities \citep[see][for a discussion on the necessity of scaling probabilities]{Kunz_2012, Jones_2018_II}. 
As a result, the free parameters in the BEAMS likelihood minimization are the $N_{\mathrm{bins}}$ offset terms $\Delta_\mu^{\BinInd}$, the nuisance parameters $\alpha$ and $\beta$, the SN Ia intrinsic scatter term $\sigma_{\mathrm{Ia, int}}$ and the scaling factor $S_{\text{non-Ia}}$. In this analysis, we use twenty logarithmically equally spaced redshift bins.

Modelling the contamination likelihood term $D_\text{non-Ia}$ (equation~\ref{eq:likelihoods}) is more difficult because core-collapse SNe are not standardized by the SALT2 framework. 
Qualitatively, we expect the distribution of non-Ia SN distance moduli to have a larger scatter and to be shifted from $\mu_{\mathrm{ref}}$ by a positive offset because non-Ia SNe are generally fainter than SNe Ia.

As BEAMS is designed to handle both SNe Ia and non SNe Ia, we do not apply a $P_{\mathrm{Ia}}$ cut prior to the BBC fit. However, in Appendix~\ref{appendix_P05}, we discuss the effects (and disadvantages) of combining BEAMS with (for example) a $P_{\mathrm{Ia}}>0.5$ selection and motivate the absence of this cut.

\subsection{Modelling the contamination likelihood}
\label{sec:cc_mu}

\begin{figure}
    \includegraphics[width=\linewidth]{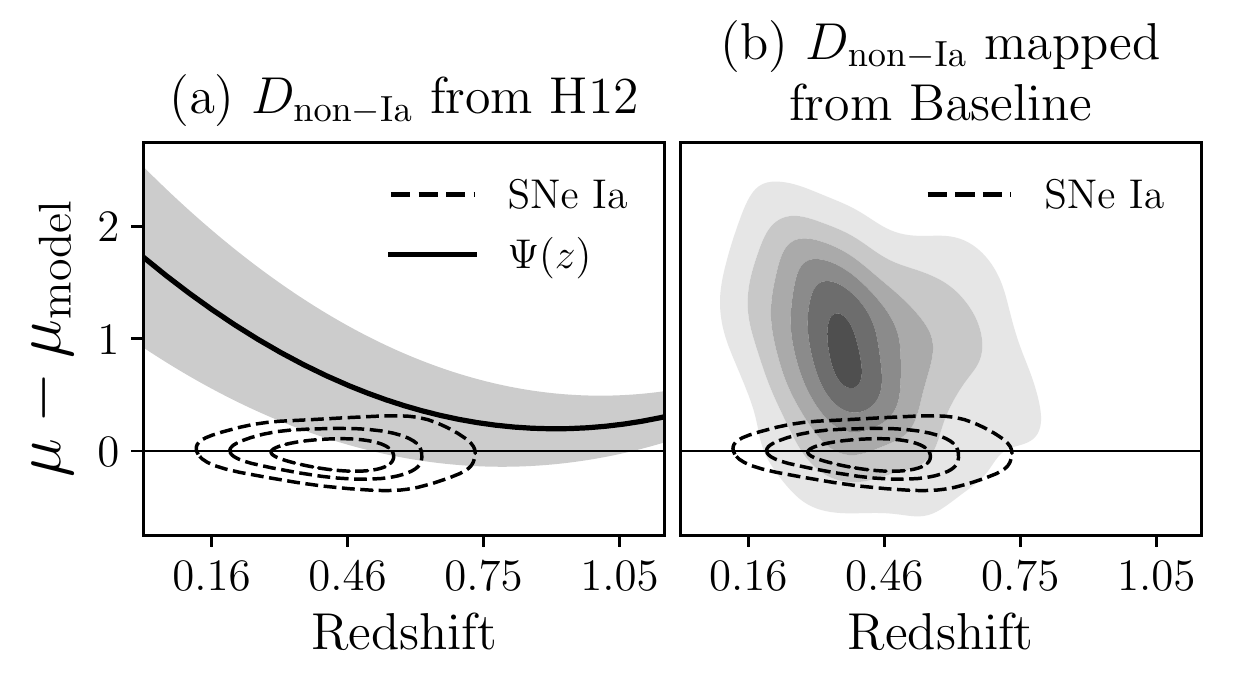}
    \caption{Modelling of core-collapse SN Hubble residuals versus redshift using the different approaches discussed in Section~\ref{sec:cc_mu}. Panel \textbf{(a)}: the modelling of Hubble residuals for the \citetalias{Hlozek_2012} approach. The black curve and grey shaded region show the best fitting polynomial $\Psi(z)$ and intrinsic scatter, $\sigma_{\text{non-Ia}}$. For comparison, we also show the Hubble residual distribution for a sample of simulated SNe Ia (dashed contours).
    Panels \textbf{(b)}: as Panel \textbf{(a)}, but when applying the approach by \citet{Kessler_2017} and using simulations.}
    \label{fig:cc_maps}
\end{figure}

We test two different approaches to describe $D_\text{non-Ia}$ analytically. The first follows \citet{Hlozek_2012}, who tested an approximation in which core-collapse SN distance moduli and intrinsic scatter are parametrized similarly to SNe Ia
\begin{equation}
  \begin{aligned}
    D_{\text{non-Ia}} &= \text{exp}  \biggl(-\frac{(\mu_{\text{obs},i} - \mu_{\text{ref, non-Ia}}(z_i))^2} {\sigma_{\mu,i}^{2}} \biggr)\\
  \end{aligned}
  \label{eq:cc_likelihoods}
\end{equation}
where
\begin{equation}
    \mu_{\text{ref, non-Ia}} = \mu_{\text{ref}}+ \Psi(z) 
    \text{ and }
    \sigma_{\text{Ia, int}} \rightarrow  \sigma_{\text{non-Ia, int}}(z),
    \label{eq:h11}
\end{equation}
and $\Psi(z)$ describes the brightness offset of the population of contaminants, and $\sigma_{\text{non-Ia, int}}$ is the redshift dependent intrinsic scatter of contaminants that is included in $\sigma_{\mu,i}$ in eq. \ref{eq:cc_likelihoods}. Both terms are modelled as second order polynomials, the coefficients of which are fitted during the BBC fit. This parametrization introduces six additional free parameters in the likelihood in equation~\ref{eq:tot_likelihoods}.
Fig.~\ref{fig:cc_maps}$a$ shows an example of the best fit $\Psi(z)$ (and relative $\sigma_{\text{non-Ia}, \text{int}}(z)$) measured from the Baseline simulations (Section~\ref{sec:simulations}).

\cite{Kessler_2017} introduced an alternative approach,
and determine the term $\mu_{\text{ref, non-Ia}}$ in equation~\ref{eq:h11} from simulation of core-collapse SNe. The mean and dispersion of the core-collapse SN distance moduli are measured from the simulation at different redshift bins. In this approach, there are no extra free parameters in the BBC fit. 

Following this approach, we use our Baseline simulation to derive the core-collapse distribution on the Hubble diagram and we show the simulated $\mu_{\text{ref, non-Ia}}$ vs. redshift in Fig.~\ref{fig:cc_maps}b.

\subsection{Bias corrections}
\label{sec:biascor}

All SN surveys are affected by selection effects introduced by their flux-limited nature. These effects introduce systematic biases in cosmological analyses of SN Ia samples, and thus SN Ia distances are corrected for such biases (equation~\ref{eq:tripp}). The corrections are generally estimated using large SN Ia Monte Carlo simulations that accurately model the survey detection efficiency and other potential selection effects \citep{1999AJ....117.1185H,2009ApJS..185...32K, 2010AJ....140..518P,Betoule_2014, DES_biascor}. Early use of simulations modelled distance bias corrections as a function of redshift only \citep{2009ApJS..185...32K, Jones_2018_II, Betoule_2014}, but \citet{2016ApJ...822L..35S} showed that this approach is not adequate because distance biases also depend on colour and stretch.

We estimate bias corrections, $\Delta\mu_\mathrm{bias}$, using the BBC framework and the simulations following Section~\ref{sec:simulations}, but including only normal SNe Ia. BBC determines an average $\Delta\mu_\mathrm{bias}$ in a five dimensional grid $\{z, x_1, c, \alpha, \beta\}$. 
For each event, the bias is interpolated between neighboring bins in the subspace of $\{z,x_1,c\}$, and also interpolated in a 2$\times$2 grid of $\alpha$ and $\beta$ ($\alpha$ in $[0.12, 0.20]$ and $\beta$ in $[2.3, 3.6]$). The simulations are used to bias correct both the real DES-SN sample and the simulated DES-SN samples. We note that bias corrections are applied prior to the BEAMS likelihood minimization presented in Section~\ref{sec:BEAMS} and they have been shown to have a weak dependence over $\alpha$ and $\beta$.

The simulations used to model bias corrections include 770,000 DES-SN events and 145,000 low-$z$ SN events (this corresponds to 500 realisations of the DES-SN sample and 500 realisations of the low-$z$ sample). The underlying assumption of BBC is that the bias correction simulation accurately describes the intrinsic properties of the SNe Ia and survey selection effects. Incomplete modelling of one of these aspects may result in inaccurate bias corrections \citep[see][for example]{DES_massstep, 2021arXiv210201776P}. The degree to which core-collapse SN contamination can affect the modelling of the SN Ia intrinsic population (and therefore bias corrections and cosmology) will be explored in future analyses. 

In the BBC approach, some cells in the five-dimensional parameter space have too few events (or no events) to reliably estimate bias corrections. SNe in these cells cannot be bias corrected and are rejected from the sample and the cosmological fit. This implicit cut further reduces the sample size, and affects SNe Ia and core-collapse SNe differently. The requirement of a valid bias correction is therefore an implicit photometric classifier for our sample. In Table~\ref{table:cuts}, we report the numbers of SNe for which a valid bias correction cannot be estimated. In the low-$z$ sample, 24 observed SNe Ia do not have valid bias corrections (approximately 8 per cent of the low redshift sample), and the simulated prediction is $18$ SNe Ia on average, in good agreement with the data. In the DES-SN samples, there are 73 SNe without valid bias corrections in the observed sample ($<4$ per cent) and the simulated prediction is $61$ SNe on average. In our simulations, we find that almost 65 per cent of the SNe without valid bias corrections are core-collapse SNe or peculiar SNe Ia, illustrating the implicit classifier in BBC. We discuss this further in Section~\ref{sec:biascorr_contam}. 

\subsection{Outlier rejection: Chauvenet's criterion}
\label{sec:outlier_rejection}

\begin{figure}
\centering
\includegraphics[width=0.85\linewidth]{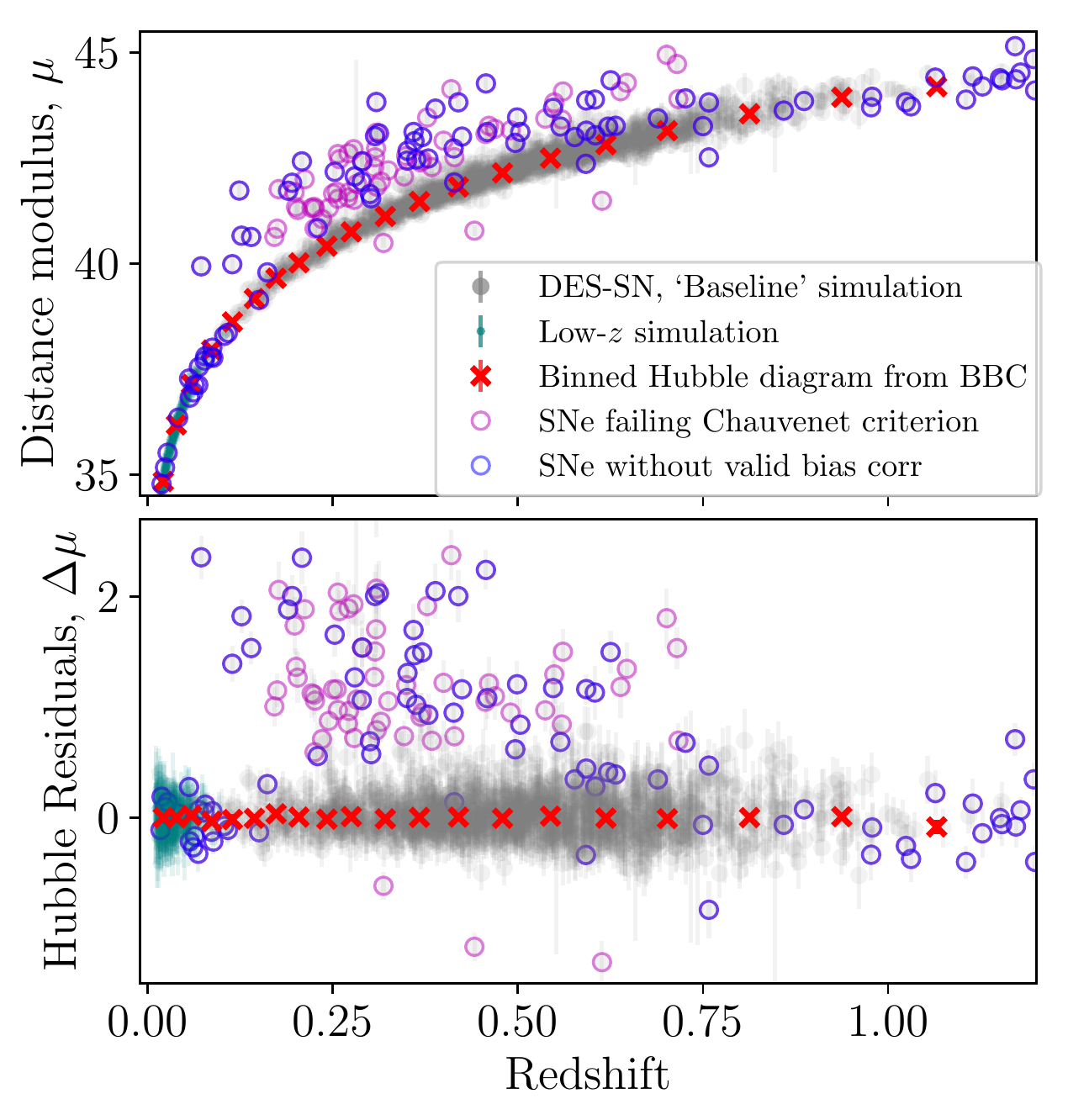}
\caption{Simulated Hubble diagram (upper panel) and Hubble residuals (lower panel) for a single realisation of the DES-SN sample (Baseline simulation, grey symbols) and the low-$z$ sample (teal symbols). We apply the SALT2 selection criteria described in Section~\ref{sec:salt2_cut}, but no other selection. SNe without a valid bias correction (Section~\ref{sec:biascor}) and/or failing Chauvenet's criterion (Section~\ref{sec:outlier_rejection}) are indicated with different colours.}
    \label{fig:example_HD}
\end{figure}

Following \citet{2011ApJS..192....1C} many cosmological analyses use Chauvenet's criterion \citep{1997ieas.book.....T} to reject outliers on the Hubble diagram \citep{Foley_Foundation, scolnic2018, DES_syst}, i.e., outliers in $\Delta\mu$. Given the number of SNe in the Hubble diagram and assuming their Hubble residuals are normally distributed around zero, Chauvenet's criterion can be used to identify the probability threshold (or $\sigma$ cut) above which the expected number of data points is below unity (i.e. less the one event is expected to have such a large deviation from zero). 

This approach has been used for samples of spectroscopically confirmed SNe Ia. In analyses of pure SNe Ia samples, Chauvenet's criterion selects normal SNe Ia and rejects atypical events or those that have poorly modelled peculiar velocities (for low redshift SNe especially).

In a photometric SN sample like the DES-SN sample, applying Chauvenet's criterion primarily rejects core-collapse SN contaminants that, in this case, are the main source of outliers in the Hubble diagram. Since we mainly focus on exploiting photometric classifiers to describe contamination (see Section~\ref{sec:photometric-classification}), rather than outlier rejection or other sigma-clipping methods, we do not apply Chauvenet's criterion by default. However, we examine the difference in cosmological parameters between using photometric classifiers and applying Chauvenet's criteria with $P_{\mathrm{Ia}}=1$ for all events (see Fig.~\ref{fig:flow_chart}). This second approach is effectively the same approach applied to analyses of spectroscopic SN sample, and it enables us to quantify cosmological biases from naively analysing a contaminated SN sample as a pure sample of spectroscopically confirmed SNe Ia.

For simplicity, we apply Chauvenet’s criterion before the BBC fit, using approximate Hubble residuals computed from initial values of the nuisance parameters ($\alpha=0.14$, $\beta=3.1$) and our reference cosmology.

For our sample of 1995 SNe following SALT2 selection (Section~\ref{sec:salt2_cut}), Chauvenet's criterion corresponds to a $4\sigma$ cut. This cut may affect the low-$z$ and DES-SN samples in different ways. For the low-$z$ sample, Chauvenet's criterion selects normal SNe Ia and rejects atypical events or those that have poorly modelled peculiar velocities. In the DES-SN sample, the criterion primarily affects core-collapse SN contaminants.
To avoid conflating the different effects of Chauvenet's criterion, we \textit{always} apply Chauvenet's criterion to the low-$z$ sample, effectively freezing these samples across our tests.

Applying Chauvenet's criterion to our observed samples removes no SNe Ia from the Foundation sample,\footnote{Chauvenet's criterion has already been applied to the Foundation DR1 sample and removes 9 SNe Ia (5 per cent of the sample). See table 7 by \citet{Foley_Foundation}.} 3 SNe Ia from the CfA+CSP samples and 122 SNe from the DES-SN sample (approximately 7 per cent of the sample). From our simulated low-$z$ samples, we predict no loss of low-$z$ SNe after applying the criterion because our low-$z$ simulation consists of normal SNe Ia without contamination. For the DES-SN sample we predict a reduction from an average of 1650 SNe to 1572 SNe (a loss of 78 SNe, approximately 5 per cent of the sample) using the \lq Baseline\rq\ simulation, in slight tension with the data.
Table~\ref{table:cuts} summarizes these numbers.

\subsection{Cosmological parameter estimation}
\label{sec:cosmo_estimation}

The output of the BBC fit is a redshift-binned Hubble diagram corrected for selection effects and contamination, and the associated diagonal covariance matrix, $C_{\mathrm{stat}}$, that includes statistical uncertainties only. As a result of the binning, the dimension of the covariance matrix is reduced from $N_{\text{SNe}}$ to $N_{\text{bins}}$. 

We note that binning the Hubble diagram may inflate systematic uncertainties that are not primarily redshift dependent \citep*{BinningSinning}. We will illustrate this uncertainty inflation for some systematics associated with SN photometric classification (Sec.~\ref{sec:cont_eff}), which may be self-calibrated in an unbinned approach. 

Finally, we estimate cosmological parameters. We test two cosmological models: a flat $w$CDM model and a flat $w_0w_a$CDM model. In both models, the dark energy equation-of-state is parametrized as $\rho \propto a^{{-3(1+w)}}$, where $\rho$ is the dark energy density and $a$ is the scale factor and it is $a=(1+z)^{-1}$; however, while a $w$CDM model assumes constant $w$, a $w_0w_a$CDM model assumes $w=w_0 + w_a (1-a)$.
Unless otherwise stated, we measure cosmological parameters assuming a prior on $\Omega_M$ of 0.311$\pm$0.010, following the cosmic microwave background measurements published by \citet{collaboration2018planck}. In future cosmological analyses of the DES photometric SN sample, SN constraints will be combined with the full CMB likelihpood from \citet{collaboration2018planck}. In Section~\ref{sec:priors} and Fig.~\ref{fig:contours}, we will show that CMB constraints constitute a more stringent prior compared to a Gaussian $\Omega_M$ prior, and thus contribute to reduce both $w$-biases due to contamination and statistical uncertainty on $w$.

When testing a flat $w$CDM model, we measure cosmological parameters using a simple $\chi^2$-minimization program that has evolved from the analysis of \citet{2011ApJS..192....1C}. This program evaluates the $\chi^2$ between $\mu_{\mathrm{Ia}}^{\BinInd}$ produced by BBC and $\mu_\mathrm{ref}$ over a grid of $\Omega_M$, $w$ and $\mathcal{M}_B$ values (assuming a flat universe) and estimate $\Omega_M$ and $w$ marginalised over $\mathcal{M}_B$ \citep[see][for a description of the $\chi^2$ definition and marginalization]{Goliath2001}. This program does not provide the full posterior distribution of the cosmological parameters we are interested to constrain. However, it is faster than most cosmological fitting programs and it is adequate for measuring biases on $w$.

To measure cosmological contours and to test a flat $w_0w_a$CDM model, we use the Cosmological Monte Carlo software CosmoMC \citep{Lewis_2002}. For the DES-SN data, absolute estimates of the cosmological parameters are blinded and only relative \textit{differences} between cosmological fits are examined.

\section{Photometric classification}
\label{sec:photometric-classification}

We use the SuperNNova \citep[SNN;][]{2020MNRAS.491.4277M} framework to perform photometric classification of our observed and simulated SN datasets, and measure for each SN event its probability of being a SN Ia, $P_{\text{Ia}}$. We choose \snn\ as the code is publicly available, and \snn\ has demonstrated good classification performance in the literature. For comparison with \snn, we also use two simple algorithms to assign $P_{\mathrm{Ia}}$:
\begin{itemize}
    \item \texttt{Perfect}: an ideal classifier, that assigns $P_{\text{Ia}}=1$ to SNe Ia and $P_{\text{Ia}}=0$ to peculiar SNe Ia or core-collapse SNe. This approach can only be used in simulations, where the true types are known;
    \item \texttt{\UnityClass}: a classifier that assigns $P_{\text{Ia}}=1$ to every SN.
\end{itemize}

\subsection{SuperNNova}
\label{sec:SNN}

\begin{table*}
\caption{Details of the different \snn\ training samples.}
\label{table:training_samples}
\begin{tabular}{ccccccc}
\hline
\snn\ & Simulation used & Core-collapse SN & Normalisation  & Number of SNe in & \multicolumn{1}{c}{Percentage of Ia, pec Ia and core-collapse}\\
 model name & for SNN training & template library& & training sample & in the training sample \\
\hline
\texttt{SNN(Base)} & Baseline & \citetalias{Vincenzi_2019} & cosmo & 287,000 & 50, 6, 44 \\
\texttt{SNN(J17)}  & J17 & \citetalias{Jones_2017_I} & cosmo & 287,000 & 50, 3, 47 \\
\texttt{SNN(DES-CC)}  & DES-CC & \citetalias{DES-CC} & cosmo &  240,000 &  50, 5, 45 \\
\texttt{SNN(global)}  & Baseline & \citetalias{Vincenzi_2019} & {global} & 287,000 & 50, 6, 44 \\
\texttt{SNN(randomHost)} & Baseline, random & \citetalias{Vincenzi_2019} & cosmo &  155,700 & 50, 5, 45\\
 & host association &  &  &  &  \\

\hline
\end{tabular}
\end{table*}

\begin{table*}
    \centering
    \caption{Contamination and efficiency measured for the \texttt{\UnityClass} classifier (rows) on different simulations (columns) after applying a $P_{\mathrm{Ia}}>0.5$ cut.}
    \label{tab:contam_table_allSNIa}
\begin{tabular}{p{4cm}>{\centering}p{1.4cm}>{\centering}p{1.2cm}>{\centering}p{1.2cm}>{\centering}p{1.2cm}>{\centering}p{1.2cm}>{\centering}p{1.2cm}>{\centering}p{0.05cm}>{\centering\arraybackslash}p{1.2cm}}

\hline
Selection criteria&  \multicolumn{6}{c}{Contamination} & & \multicolumn{1}{c}{Efficiency} \\
&  Only pec Ia &  Baseline &  LFs+Offset &  Dust(H98) &  J17 &  DES-CC & & (Baseline) \\
\hline
\texttt{\UnityClass}, no SALT2 selection $^\dagger$ & \gradient{2.6} & \multicolumn{1}{c}{\gradient{22.5}} & \gradient{31.7} & \gradient{22.0} &  \gradient{28.5} &  \gradient{25.8} & & - \\
\texttt{\UnityClass} & \gradient{2.1} & \multicolumn{1}{c}{\gradient{8.2}} & \gradient{11.6} & \gradient{8.5} &  \gradient{8.7} &  \gradient{9.8} & & \NDgradient{100.0} \\
\texttt{\UnityClass+Chauvenet} & \gradient{1.0} & \multicolumn{1}{c}{\gradient{3.1}} & \gradient{5.3} & \gradient{3.4} &  \gradient{3.7} &   \gradient{3.2} & & \NDgradient{98.7} \\
\texttt{\UnityClass+Chauvenet}, $c<$0.15    & \gradient{0.7} & \multicolumn{1}{c}{\gradient{2.2}} & \gradient{4.0} & \gradient{2.3} &  \gradient{1.6} &   \gradient{2.5} & & \NDgradient{89.4} \\
\hline
\end{tabular}
    \begin{tablenotes}\footnotesize
        \item $^\dagger$ Fraction of contaminants after SALT2 fit loose cuts of $x_1 \in [-4.9,4.9]$ and $c \in [-0.49,0.49]$ (i.e., without applying the SALT2-based selection discussed in Section~\ref{sec:salt2_cut}). \citepalias[see][]{Vincenzi_2020}.
    \end{tablenotes}
\end{table*}

\begin{table*}
    \centering
    \caption{Contamination and efficiency measured for different SNN models (rows) tested on different simulations (columns).}
    \label{tab:contam_table}

\begin{tabular}{p{2.2cm}>{\centering}p{1.4cm}>{\centering}p{1.2cm}>{\centering}p{1.2cm}>{\centering}p{1.2cm}>{\centering}p{1.2cm}>{\centering}p{1.2cm}>{\centering}p{0.05cm}>{\centering\arraybackslash}p{1.2cm}}

\hline
SNN model$^a$ &  \multicolumn{6}{c}{Contamination after \emph{testing} SNN on different simulations} & & \multicolumn{1}{c}{Efficiency} \\
 &  Only pec Ia &  Baseline &  LFs+Offset &  Dust(H98) &  J17 &  DES-CC & & (Baseline) \\
\hline
\texttt{SNN(Base)}  & \gradient{0.4} &  \multicolumn{1}{c}{\textbf{\gradient{0.8}}$^b$} & \gradient{1.1} & \gradient{0.9} & \gradient{1.0} &  \gradient{1.4} & & \NDgradient{99.5} \\ 
\texttt{SNN(J17)}   & \gradient{0.7} & \multicolumn{1}{c}{\gradient{1.7}} & \gradient{2.8} & \gradient{1.9} &  \textbf{\gradient{1.0}$^b$} &   \gradient{2.1} & & \NDgradient{99.2} \\ 
\texttt{SNN(DES-CC)}  & \gradient{0.9} & \multicolumn{1}{c}{\gradient{2.0}} & \gradient{3.2} & \gradient{2.3} &  \gradient{1.9} &   \textbf{\gradient{1.6}$^b$} & & \NDgradient{99.0} \\
\texttt{SNN(global)} & \gradient{0.8} & \multicolumn{1}{c}{\gradient{2.1}} & \gradient{3.5} & \gradient{2.1} &  \gradient{1.4} &   \gradient{2.3} & & \NDgradient{97.7} \\
\texttt{SNN(randomHost)} & \gradient{0.7} & \multicolumn{1}{c}{\gradient{1.3}} & \gradient{1.9} & \gradient{1.5} &  \gradient{1.3} &   \gradient{1.6} & & \NDgradient{98.1} \\
\hline
\end{tabular}
    \begin{tablenotes}\footnotesize
        \item $^{a}$ See Table~\ref{table:training_samples} for a description of the training approach utilised for each SNN model.
        \item $^{b}$ We highlight in bold the contamination measured using the same simulation both for training and testing.
    \end{tablenotes}
\end{table*}

\snn\ is an open-source\footnote{\url{https://github.com/supernnova/SuperNNova}} machine learning algorithm that implements Recurrent Neural Networks for photometric classification of SNe. It is trained to classify different types of transients using photometric data only (i.e., fluxes and flux uncertainties in different filters) and, optionally, redshift information. It does not rely on feature extraction or light-curve fitting.

Several metrics can be used to assess the performance of \snn. In the binary classification method, these are based on the number of true positives (TPs; SNe Ia correctly classified as such), true negatives (TNs; core-collapse SNe correctly classified as such), false positives (FPs; core-collapse SNe incorrectly identified as SNe Ia) and false negatives (FNs; SNe Ia identified as core-collapse). Following \citet{2020MNRAS.491.4277M}, the contamination (by core-collapse SNe, or peculiar SNe Ia) of the classified photometric SN Ia sample and the classification efficiency are defined as
\begin{equation}
\label{eq:contamination}
\text{Contamination}=\frac{\text{FP}}{\text{FP}+\text{TP}}
\end{equation}
and 
\begin{equation}
\label{eq:efficiency}
\text{Efficiency}=\frac{\text{TP}}{\text{TP}+\text{FN}}.
\end{equation}
We implement \snn\ using the same hyper-parameters as \citet{2020MNRAS.491.4277M}, and include spectroscopic redshift information. 

For our analysis, we normalise the input fluxes using the \lq cosmo\rq\ method (Moller et al. in prep.). In this method, each SN multi-band light curve is normalised independently and the normalization factor is the SN maximum flux (in any filter). This method makes \snn\ agnostic to the relative differences in apparent brightness between SNe, while preserving colour and signal-to-noise information (flux uncertainties are normalised using the same factor as for fluxes). With this normalisation, rescaled fluxes close to zero correspond to early/late data points and rescaled fluxes close to one correspond to data points around peak brightness.

We also test an alternative normalization method labelled as \lq global\rq. In this method, the normalisation factors are estimated from the full sample of light curves and the same normalisation is applied to all light curves. This method preserves the relative brightnesses between different SNe and the full range of magnitudes. As a result, the brightest (lower redshift) SNe have rescaled fluxes closer to one, while faintest SNe have rescaled fluxes closer to zero.

\subsection{Training of \snn}
\label{sec:SNNtraining}

\snn\ requires training on very large samples of SNe ($>$100,000 events). Combining all SN surveys from the last 15 years, the sample of spectroscopically-confirmed SNe available is around 10,000 events\footnote{Source: Transient Name Server, \url{https://wis-tns.weizmann.ac.il/}}; it is an inhomogeneous sample with an uncertain selection function and biased towards bright, lower-redshift events. To obtain a training sample with sufficient statistics, \snn\ relies on large simulations where the SN Ia and SN non-Ia rest-frame SED models are derived from spectroscopically confirmed events.

To generate the training samples we combine 100 realisations of our Baseline simulation, apply a simple selection to the simulated events \citep[at least two detections, applying the detection efficiency presented by][]{2015AJ....150..172K}, and apply the host galaxy spectroscopic efficiency of  \citetalias{Vincenzi_2020}. We do not apply any additional spectroscopic classification efficiency like the one applied to the training samples generated for the SN classification challenges presented by \citet{2010arXiv1001.5210K,2018arXiv181000001T}. Moreover, we do not perform SALT2 fits for \snn.
We also generate three additional training samples, using the J17 simulation (\texttt{SNN(J17)}), the DES-CC simulation (\texttt{SNN(DES-CC)}), and the Baseline simulation with host galaxies assigned randomly (\texttt{SNN(randomHost)}).

To compare the two different normalizations in \snn, we also train a model using the Baseline simulation and the global normalisation method instead of the cosmo normalisation (\texttt{SNN(global)}). This tests the effects of a classifier that has knowledge of the relative brightnesses between SNe Ia and core-collapse SNe. A summary of the five \snn\ models and the assumptions in their training simulations is in Table~\ref{table:training_samples}.

\subsection{Contamination and Efficiency}
\label{sec:cont_eff}

\begin{figure*}
    \includegraphics[width=0.9\linewidth]{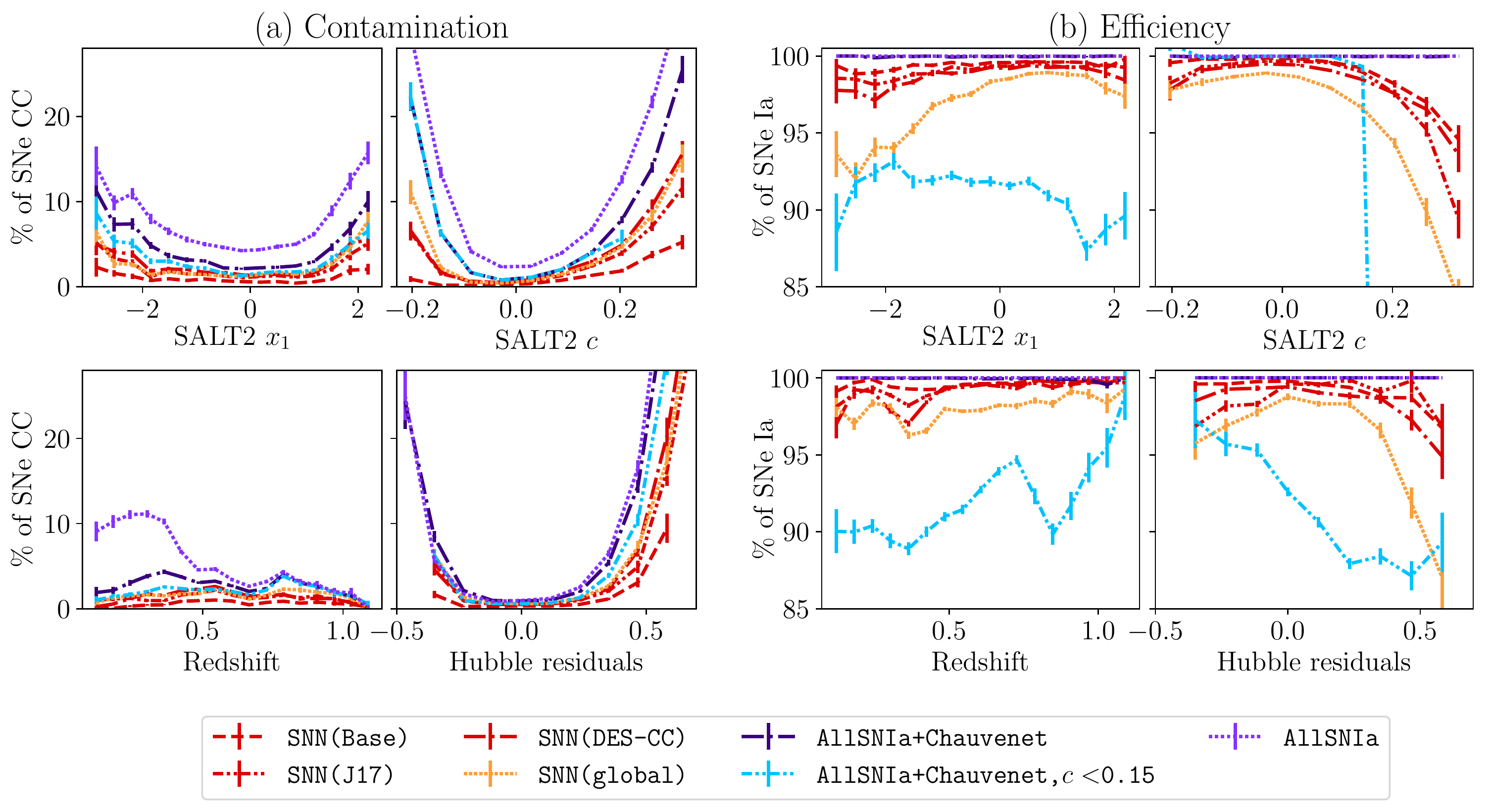}
    \caption{Contamination (panel \textbf{(a)}) and efficiency (panel \textbf{(b)}) using three \snn\ models \texttt{SNN(Base)}, \texttt{SNN(J17)} and \texttt{SNN(DES-CC)} measured on our Baseline simulation. All contamination and efficiency percentages are measured relative to the bin, \emph{not} relative to the total sample. In panel \textbf{(a)}, we present contamination as a function of SALT2 $x_1$ (upper left), $c$ (upper right), redshift (lower left) and Hubble residual (lower right). Panel \textbf{(b)} is the same as panel \textbf{(a)}, but showing efficiency. Contamination and efficiency are defined in Section~\ref{sec:SNN}.}
    \label{fig:contam}
\end{figure*}

\begin{table}
\caption{Fraction of different sub-types of contaminants for different selection cuts. Contamination is measured on the Baseline simulation, after applying the SALT2-based selection described in Section~\ref{sec:salt2_cut} and Chauvenet's criterion. All contamination percentages are measured relative to the bin, \emph{not} relative to the total sample.}
\label{tab:contam_table_type}
\begin{tabular}{lcccc}
\hline
{Selection} &  \%non-Ia SNe & \% Pec Ia &  \% II &  \% Ibc \\
\midrule

$c>0.2$    &      16.6 &        5.3 &    0.6 &    10.7 \\
$c<-0.2$   &      24.1 &        0.1 &   22.2 &     1.8 \\
$x_1>2$   &      12.0 &        2.3 &    2.8 &     6.9 \\
$x_1<-2$  &        6.9 &        0.7 &    1.6 &     4.6 \\
$\mathrm{log}_{10}M_{\star}/M_{\odot}<10$ &   2.8 &        0.8 &    0.8 &     1.1 \\
$\mathrm{log}_{10}M_{\star}/M_{\odot}>10$ &    3.6 &        1.5 &    1.0 &     1.1 \\
\hline
\end{tabular}
\end{table}

We test \snn\ on the simulations summarised in Section~\ref{sec:simulations}, measuring the average contamination and efficiency after our standard selection (Section~\ref{sec:salt2_cut}) and after requiring $P_{\text{Ia}}>0.5$ cut. As already mentioned in Section~\ref{sec:BEAMS}, BBC is designed to handle both SNe Ia and non SNe Ia, therefore we do not require a $P_{\text{Ia}}>0.5$ cut in the cosmological sample (see Appendix \ref{appendix_P05}). 

We first examine the case of no classifier (i.e., \texttt{\UnityClass}) in Table~\ref{tab:contam_table_allSNIa}) and SALT2-based selection. Applying only SALT2-based selection reduces contamination to less than 12 per cent, a factor of two smaller compared to SN samples before SALT2-based selection. When combined with outlier rejection (\texttt{\UnityClass+Chauvenet}, see Section~\ref{sec:outlier_rejection}), the contamination reduces to 4.0--6.6 per cent. A tighter SALT2 colour selection (Section~\ref{sec:salt2_cut}) combined with Chauvenet's criterion ( \texttt{\UnityClass+Chauvenet,$c<0.15$}), reduces the contamination further to 1.6--4.0 per cent. These results set a level of comparison for assessing the performance of \snn.

The performance of the \snn\ models is shown in Table~\ref{tab:contam_table}. For the \snn\ models \texttt{SNN(Base)}, \texttt{SNN(J17)} and \texttt{SNN(DES-CC)}, the performance is improved compared to outlier rejection methods only, with contamination of 0.8--3.2 per cent and an efficiency equal or above 99 per cent. 
\texttt{SNN(Base)}, trained on our Baseline simulation, performs well not only when tested on Baseline simulations (0.8 per cent contamination), but also when tested on the simulations J17 and DES-CC, with contamination of 1.0 and 1.4 per cent respectively. In these two cases, the \texttt{SNN(Base)} classifier is trained on core-collapse SN templates that are independent from the ones used to generate the simulations, suggesting that the \texttt{SNN(Base)} model generalizes well.

By contrast, the \texttt{SNN(J17)} and \texttt{SNN(DES-CC)} classifiers perform well when tested on simulations generated using the same core-collapse SN models (in bold in Table~\ref{tab:contam_table}), but when tested on Baseline simulations they predict levels of contamination that are two and three times larger compared to using the \texttt{SNN(Base)} model. This difference reflects the increased diversity of contaminants in the Baseline simulation compared to the J17 and DES-CC simulations.
    
We make two further observations. The first is that, following the application of \snn, peculiar SNe Ia account for around a third to a half of the contamination (Table~\ref{tab:contam_table}), suggesting that this class of transients plays an important role in our analysis, and that they are as difficult to identify as core-collapse SNe with the current training set and configuration.\footnote{To improve classification of peculiar SNe Ia, the fraction of this sub-type of SNe could be augmented in the training set.} The second is that, comparing the Baseline and Dust(H98) simulations, we do not observe large differences in the contamination even though none of the \snn\ models have been trained using the full range of dust extinction included in the Dust(H98) simulation. This result suggests that including dust extinction in the simulations that is unmodelled in the training samples does not significantly affect classification performance.

\subsubsection{Performance as a function of SN Ia properties}

Fig.~\ref{fig:contam} shows the contamination and efficiency for the Baseline simulation as a function of redshift, fitted $x_1$ and $c$, and $\Delta\mu$. These plots identify regions of parameter space where non-Ia SN contamination is higher (or efficiency is lower). The poorest performance in terms of contamination \emph{per-bin} is observed at the extremes of the SALT2 parameter distributions.

Focusing on SALT2 $c$, contamination increases significantly for very blue events ($>20$ per cent for $c<-0.2$), mainly due to fast-declining type II and type IIn SNe that are generally bluer than SNe Ia at peak. Similarly, classification is more difficult for redder SNe ($>10$ per cent contamination and $<95$ per cent efficiency for $c>0.2$), where intrinsically redder and lower signal-to-noise stripped envelope SNe are more easily misclassified as red (and therefore also faint) SNe Ia, and vice versa (see Table~\ref{tab:contam_table_type}). Contamination is less than 2 per cent for  $-0.1<c< 0.1$, even when only applying the \texttt{\UnityClass} classifier and Chauvenet's criterion.
For stretch, contamination at higher $x_1$ values is mainly due to slower declining stripped-envelope SNe, while contamination at the low $x_1$ is dominated by faster declining SNe Ic (see Table~\ref{tab:contam_table_type}).

\subsubsection{Performance of Global versus Cosmo normalisations}
\label{sec:cont_eff_globalcosmo}
Contamination after using \snn\ models trained with the global SNN normalisation (\texttt{SNN(global)}) is similar to the other \snn\ models trained using the cosmo normalisation. However, \texttt{SNN(global)} has a significantly lower efficiency -- less than 98.5 per cent -- and it decreases significantly for positive Hubble residuals.

In the \texttt{SNN(global)} model, the relative brightness between SNe Ia and core-collapse SNe is preserved both in the training and testing phase. Our results show that encoding SN relative brightnesses in the classification does not result in a significant decrease in contamination, and mainly affects the classification of faint SNe Ia. Approximately 10 to 15 per cent of SNe Ia in the faint tail of the Hubble residual distribution ($\Delta\mu>0.25$\,mag) are misclassified as non-SNe Ia.  

\subsubsection{Performance as a function of host galaxy properties}
\label{sec:cont_massstep}

\begin{figure}
\centering
\includegraphics[width=0.99\linewidth]{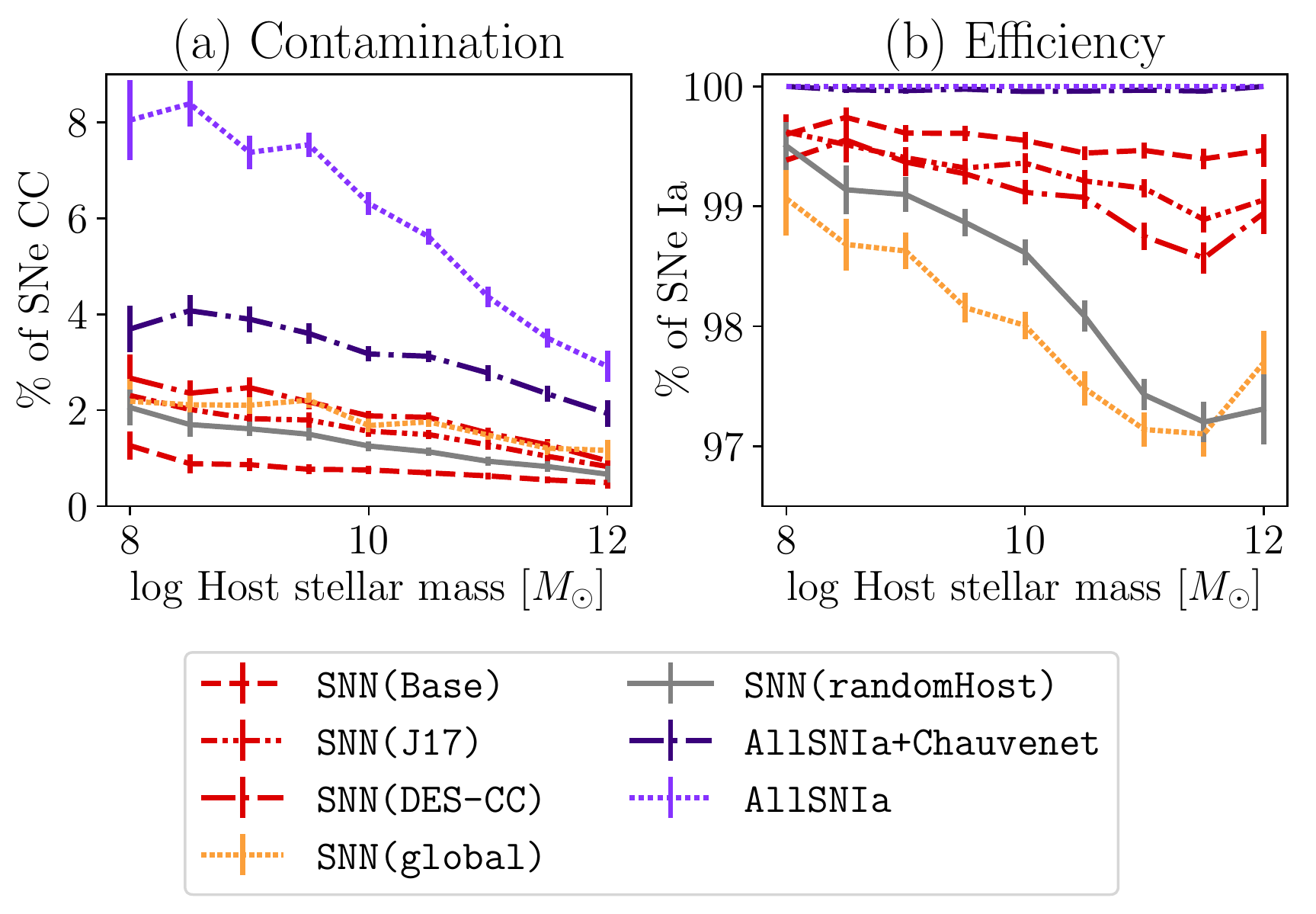}
\caption{Contamination (left panel) and efficiency (right panel) as a function of SN host-galaxy stellar mass. We measure contamination and efficiency for different \snn\ models (see Section~\ref{sec:SNNtraining}) and before/after applying the Chauvenet's criterion (see Section~\ref{sec:outlier_rejection}).}
    \label{fig:contam_massstep}
\end{figure}

\begin{table*}
    \centering
    \caption{As Table~\ref{tab:contam_table_allSNIa}, but following the application of BBC bias corrections.}
    \label{tab:contam_table_allSNIa_postBBC}
\begin{tabular}{p{4cm}>{\centering}p{1.4cm}>{\centering}p{1.2cm}>{\centering}p{1.2cm}>{\centering}p{1.2cm}>{\centering}p{1.2cm}>{\centering}p{1.2cm}>{\centering}p{0.05cm}>{\centering\arraybackslash}p{1.2cm}}

\hline
{Selection criteria} &  \multicolumn{6}{c}{Contamination after BBC} & & \multicolumn{1}{c}{Efficiency} \\
 &  Only pec Ia &  Baseline &  LFs+Offset &  Dust(H98) &  J17 &  DES-CC & & (Baseline) \\
\hline
\texttt{\UnityClass} & \gradient{1.9} & \multicolumn{1}{c}{\gradient{5.7}} & \gradient{8.0} & \gradient{6.4} &  \gradient{6.4} & \gradient{7.2} & & \NDgradient{100.0} \\
\texttt{\UnityClass +Chauvenet} & \gradient{1.1} & \multicolumn{1}{c}{\gradient{3.1}} & \gradient{5.0} & \gradient{3.6} &  \gradient{3.5} & \gradient{3.2} & & \NDgradient{100.0} \\
\texttt{\UnityClass +Chauvenet}, $c<0.15$ & \gradient{0.8} & \multicolumn{1}{c}{\gradient{2.2}} & \gradient{3.8} & \gradient{2.6} &  \gradient{1.8} &   \gradient{2.6} & & \NDgradient{91.5} \\
\hline
\end{tabular}
\end{table*}

\begin{table*}
    \centering
    \caption{As Table~\ref{tab:contam_table}, but following the application of BBC bias corrections.}
    \label{tab:contam_table_postBBC}
\begin{tabular}{p{2.2cm}>{\centering}p{1.4cm}>{\centering}p{1.2cm}>{\centering}p{1.2cm}>{\centering}p{1.2cm}>{\centering}p{1.2cm}>{\centering}p{1.2cm}>{\centering}p{0.05cm}>{\centering\arraybackslash}p{1.2cm}}
\hline
SNN model$^{a}$ &  \multicolumn{6}{c}{Contamination after testing SNN on different simulations} & & \multicolumn{1}{c}{Efficiency} \\
&  Only pec Ia &  Baseline &  LFs+Offset &  Dust(H98) &  J17 &  DES-CC & & (Baseline) \\
\hline
\texttt{SNN(Base)} & \gradient{0.4} & \multicolumn{1}{c}{\textbf{\gradient{0.7}}$^b$} & \gradient{1.0} & \gradient{0.9} &  \gradient{0.9} & \gradient{1.3} & & \NDgradient{99.5} \\
\texttt{SNN(J17)} & \gradient{0.6} & \multicolumn{1}{c}{\gradient{1.5}} & \gradient{2.4} & \gradient{1.7} &  \textbf{\gradient{1.0}}$^b$ & \gradient{2.0} & & \NDgradient{99.2} \\
\texttt{SNN(DES-CC)} & \gradient{0.9} & \multicolumn{1}{c}{\gradient{1.8}} & \gradient{2.9} & \gradient{2.1} &  \gradient{1.7} & \textbf{\gradient{1.5}}$^b$ & & \NDgradient{99.0} \\
\texttt{SNN(global)} & \gradient{0.8} & \multicolumn{1}{c}{\gradient{1.7}} & \gradient{2.8} & \gradient{1.9} &  \gradient{1.3} & \gradient{2.0} & & \NDgradient{97.8} \\
\texttt{SNN(randomHost)} & \gradient{0.7} & \multicolumn{1}{c}{\gradient{1.2}} & \gradient{1.8} & \gradient{1.4} &  \gradient{1.2} & \gradient{1.6} & & \NDgradient{98.1} \\
\hline
\end{tabular}
    \begin{tablenotes}\footnotesize
        \item $^{a}$ See Table~\ref{table:training_samples} for a description of the training approach utilised for each SNN model.
        \item $^{b}$ We highlight in bold the contamination measured using the same simulation both for training and testing.
    \end{tablenotes}
\end{table*}

Our simulations are designed to account for the differing properties and rates of SNe in different host galaxies. This allows us to predict contamination in our photometric SN Ia samples as a function of host galaxy properties. As a reference, the \texttt{SNN(randomHost)} does not use these intrinsic rates and assigns host galaxies randomly.

In Fig.~\ref{fig:contam_massstep}, we present contamination and efficiency as a function of host galaxy stellar mass before applying any classification algorithm (i.e., applying only the \texttt{\UnityClass} classifier and Chauvenet's criterion) and after applying  \snn. Contamination is not equally distributed across host galaxies of different mass, but is always larger in lower mass galaxies. This variation is expected as most of the hosts in the highest mass bin consists of more passive galaxies, with a preference towards SNe Ia and only small numbers of core-collapse SNe. Therefore, the fraction of contamination in these environments is low (less than 2 per cent) even with no photometric classification.

The efficiency of classification is mostly insensitive to host galaxy stellar mass, with two exceptions: efficiencies of the models \texttt{SNN(global)} and \texttt{SNN(randomHost)} drop significantly in higher mass galaxies. For the \texttt{SNN(global)} model using the \lq global\rq\ normalisation (Section~\ref{sec:SNNtraining}), the training retains information about the relative brightnesses between SNe Ia and SN contaminants. This model is likely to heavily \lq fit\rq\ on the information that core-collapse SNe are generally fainter than SNe Ia. This means that faint SNe Ia (i.e., SNe Ia with positive Hubble residuals, see Fig.~\ref{fig:contam}b) in massive hosts (with lower signal-to-noise due to a brighter host galaxy background) are more easily misclassified as core-collapse SNe.

The \texttt{SNN(randomHost)} model is trained on a set of SN Ia light curves that have been assigned randomly to host galaxies. \citetalias{Vincenzi_2020} demonstrated that the random association of host galaxies to simulated SNe produces a distribution of host brightnesses and masses in disagreement with the data (fig.~9 in \citetalias{Vincenzi_2020}). Therefore, host galaxies in the training sample of \texttt{SNN(randomHost)} are on average fainter than those in the DES-SN sample or simulations. When the \texttt{SNN(randomHost)} model is tested on realistic SN samples, a significant fraction of SNe Ia in bright and high mass galaxies is misclassified as core-collapse SNe. This test demonstrates the importance of training machine learning algorithms like SNN on simulations that include a realistic SN-host association. Sub-populations of SNe Ia (e.g., SNe in bright galaxies) can be reduced or removed by classification simply because they are not modelled in the training sample, with a potential impact on studies of SN Ia populations and on SN Ia cosmology in general.

\begin{figure}
\centering
\includegraphics[width=\linewidth]{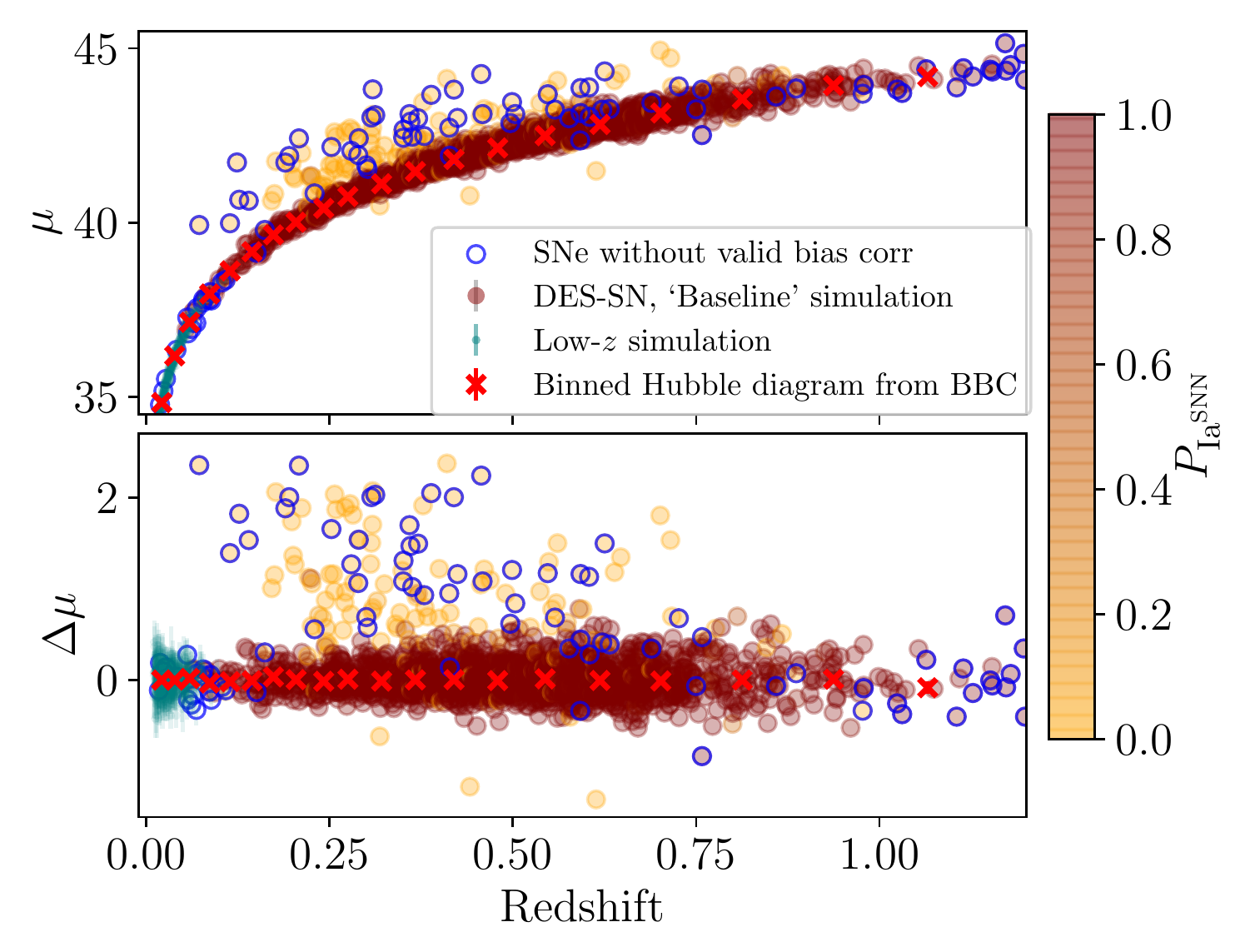}
\caption{Same as Fig.~\ref{fig:example_HD}, but here we highlight for each simulated event, its probability of being a type SN as estimated with \texttt{SNN(Base)}.}
    \label{fig:example_HD_SNN}
\end{figure}

Similarly to Fig.~\ref{fig:example_HD}, we show the Hubble diagram for a simulated sample of SNe in Fig.~\ref{fig:example_HD_SNN} and we highlight SN probabilities, $P_{\mathrm{Ia}}$, estimated applying \texttt{SNN(Base)}. 

\subsection{Effects of BBC bias corrections on contamination}
\label{sec:biascorr_contam}

In the BBC framework, there are cells of the three-dimensional sub-space $\{z, x_1, c\}$ that have no SN Ia (or too few events). Real events in those cells are rejected prior to the BBC fit and this systematically disfavours SNe that
lie in regions that are atypical for SNe Ia. As a result, the BBC bias corrections naturally reduce contamination from peculiar SNe Ia and core-collapse SNe.
Tables~\ref{tab:contam_table_allSNIa_postBBC} and~\ref{tab:contam_table_postBBC} presents contamination and efficiency after BBC bias corrections are applied (cf. Tables~\ref{tab:contam_table_allSNIa} and~\ref{tab:contam_table}, the contamination and efficiency before BBC). As expected, the number of SNe Ia is reduced by less than 1 per cent, while the number of core-collapse SNe is reduced by 20--30 per cent.

When analysing contamination after a $P_{\mathrm{Ia}}>0.5$ cut from \snn, the effect of bias corrections on the contamination is almost negligible because \snn\ is very efficient at removing contamination. However, when using no classifier (i.e., \texttt{\UnityClass}; Table~\ref{tab:contam_table_allSNIa_postBBC}) the bias corrections have a larger impact on reducing contamination. In Appendix~\ref{sec:appendix_novalidbias}, we consider the sub-sample of events that are rejected from the sample only due to the lack of a valid bias correction, and investigate the impact of including these events in the analysis by fixing their bias correction to zero.

\subsection{Comparison with the data}
\label{sec:comp_data}

\begin{figure*}
\centering
\includegraphics[width=0.9\linewidth]{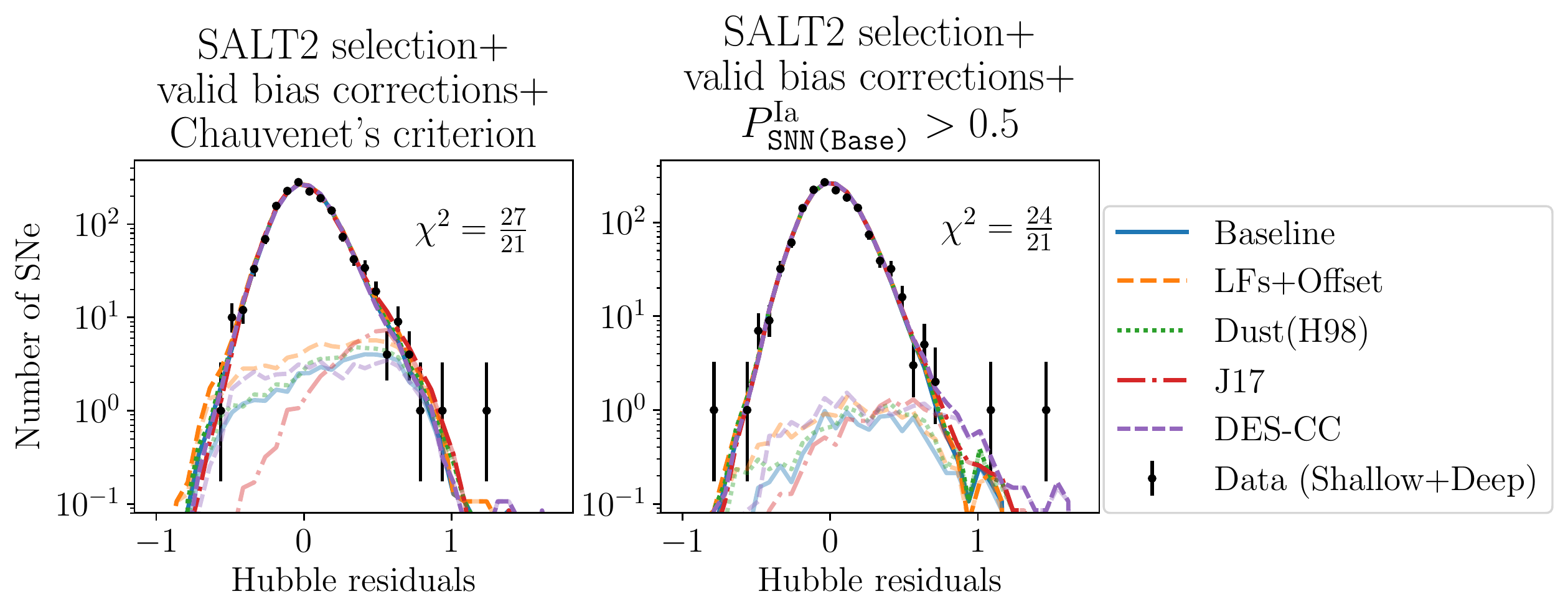}
\caption{Distributions of observed and simulated Hubble residuals for different selection criteria. Distributions are presented for the data (including Poisson uncertainties, black symbols) and for the 5 simulations summarised in Table~\ref{table:sims}: both SNe~Ia and core-collapse SNe are combined in the darker lines, and simulated core-collapse SNe only are shown in the partially transparent lines. 
\emph{Left:} Sample selected applying SALT2-based selection discussed in Section~\ref{sec:salt2_cut}, Chauvenet's criterion discussed in Section~\ref{sec:outlier_rejection} and requiring a valid bias correction (Section~\ref{sec:biascor}). \emph{Right:} Sample selected applying SALT2-based selection discussed in Section~\ref{sec:salt2_cut}, a probability cut $P_{\mathrm{Ia}}>0.5$ (where $P_{\mathrm{Ia}}$ are determined using the \snn\ classifier trained on the Baseline simulation, \texttt{SNN(Base)}) and requiring a valid bias correction. In each panel, we report the reduced $\chi^2$ between data and simulations.}
    \label{fig:comparison}
\end{figure*}

We apply bias corrections, Chauvenet's criterion and the \snn\ classifier to the DES photometric SN sample.
In Fig.~\ref{fig:comparison}, we compare the results obtained from data and from simulations for different sets of selection cuts.\footnote{A version of the same comparison \textit{before} classification-based cuts is available in \citetalias[][figure 13]{Vincenzi_2020}}

First, we consider Hubble residuals measured after applying SALT2-based selection (Section~\ref{sec:salt2_cut}), the Chauvenet's criterion (Section~\ref{sec:outlier_rejection}) and requiring a valid bias correction (Section~\ref{sec:biascor}). Simulations and data are in very good agreement (Fig.~\ref{fig:comparison}a); the asymmetry in the Hubble residual distribution due to the small fraction of core-collapse contamination ($<3.8$ per cent, see Table~\ref{tab:contam_table_allSNIa_postBBC}) is well reproduced by simulations and the reduced $\chi^2$ between data and the Baseline simulation is approximately $1.1$.

Second, we repeat the test above and additionally require $P_{\mathrm{Ia}}>0.5$, where $P_{\mathrm{Ia}}$ is estimated from the \snn classifier trained on the Baseline simulation (\texttt{SNN(Base)}). The agreement between data and simulations is also good (reduced $\chi^2$ of $0.7$) and the tail of SNe with faint Hubble residuals is significantly reduced both in the data and in the simulations (Fig.~\ref{fig:comparison}b).

We note the presence of a few outliers (Hubble residuals larger than 1 mag) in the observed Hubble residuals distribution, that are not reproduced in the simulations. This could be due to a small fraction of SNe in the DES-SN sample \citep[less than 1.1 per cent according to][]{DES_deepstacks} that is mismatched to a closer and brighter galaxy, and thus appear as faint outliers on the Hubble diagram. We remind the reader that host mismatch is not included in our simulations.

\section{Biases on cosmological parameters}
\label{sec:cosmo_bias}

The BBC framework requires several modelling choices, each causing a potential bias on the binned SN Ia distance moduli, $\mu_{\mathrm{Ia}}^{\BinInd}$, and on the resulting fitted cosmological parameters. We explore these choices in this section. The BBC configurations we test are listed in Table~\ref{table:BBC_opts} and illustrated in Fig.~\ref{fig:flow_chart}. Each is a different combination of classifier and $\mathcal{L}_{\text{non-Ia}}$. Specifically, we test:
\begin{itemize}
    \item $P_{\text{Ia}}$ measured from the five different \snn\ classifiers (Table~\ref{table:training_samples}), as well as the \texttt{Perfect} and \texttt{\UnityClass} approaches;
    \item Two approaches for the modelling of $\mathcal{L}_{\text{non-Ia}}$ (Section~\ref{sec:cc_mu}): the polynomial fitting method of \citetalias{Hlozek_2012} (\texttt{D$_{\mathrm{non-Ia}}$(H12)}), and the \citet{Kessler_2017} method implemented using the Baseline simulation (\texttt{D$_{\mathrm{non-Ia}}$(Base)}).
\end{itemize}
We test combining Chauvenet's criterion (Section~\ref{sec:outlier_rejection}) with the \texttt{\UnityClass} approach. 

We consider as our reference the configuration that uses the classifier \texttt{SNN(Base)}, and for which the core collapse SN likelihood is modelled from the Baseline simulation. This has the label \lq \texttt{SNN(Base)\,D$_{\mathrm{non-Ia}}$(Base)}\rq, and is used as the benchmark to evaluate other BBC configurations.

All our tests are run on the simulations presented in Section~\ref{sec:simulations}, reproducing the realistic scenario of testing classifiers on samples of light curves that are not in the samples used to train the classifier. This allows a verification that our modelling of $\mathcal{L}_{\text{CC}}$ is sufficiently generalised to be applied to any population of core-collapse SN contaminants. Both are critical to robustly validate our results.  

For each simulation, we estimate different cosmology-related parameters averaged over 50 realizations: $\mu_{\mathrm{Ia}}^{\BinInd}$, nuisance parameters ($\alpha$, $\beta$, $\sigma_{\mathrm{Ia, int}}$, $S_{\mathrm{non-Ia}}$), $w$, and the time-varying dark energy equation-of-state parameters $w_0$ and $w_a$. We then calculate biases due to contamination as:
\begin{equation}
    \Delta X = \langle X_{\mathrm{Ia+CC}} - X_{\mathrm{Ia\ only,\ perfect\ classification}} \rangle_{\mathrm{(50\ realizations)}}
    \label{eq:bias_definition}
\end{equation}
where $X$ represents either $\mu_{\mathrm{Ia}}^{\BinInd}$ or the nuisance parameters or cosmological parameters $w$, $w_0$, $w_a$ depending on the context. Essentially, we define a bias $\Delta X$ on a cosmological parameter $X$ due to contamination as the average difference between the value of the parameter fitted including contamination, and the value of the parameter fitted with no contamination and assuming a perfect classification. Uncertainties on $\Delta X $ are estimated as standard errors on the mean.

\begin{table*}
\caption{Summary of BBC configurations (see also Fig.~\ref{fig:flow_chart}). The second line (highlighted) lists the reference configuration.}
\label{table:BBC_opts}
\begin{tabular}{rccccc}
\hline
\multicolumn{2}{c}{BBC} & Classifier & Modelling & $\Delta w$ using & $\Delta w$ using\\
\multicolumn{2}{c}{configuration$^\textrm{(a)}$} & & of $D_{\mathrm{non-Ia}}$ & Baseline simulation$^\textrm{(b)}$ & DES-SN Data$^\textrm{(c)}$\\
\hline
1) & \texttt{Perfect  D$_{\mathrm{non-Ia}}$(Base) } & Perfect & Baseline & 0.0001$\pm$0.0002 & -\\
\rowcolor{Gray}
2)$^{*}$ & {\texttt{SNN(Base)D$_{\mathrm{non-Ia}}$(Base)}} & {SNN(Base)}  & {Baseline} & 0.0045$\pm$0.0008 & 0.0000 (0.0338) \\
3) & \texttt{SNN(J17) D$_{\mathrm{non-Ia}}$(Base) } & SNN(J17) & Baseline &  0.0109$\pm$0.0009 & 0.0059 (0.0342) \\
4) & \texttt{SNN(DES-CC)   D$_{\mathrm{non-Ia}}$(Base) } & SNN(DES-CC) & Baseline & 0.0045$\pm$0.0008 & 0.0101 (0.0324)  \\
5) & \texttt{SNN(Base)   D$_{\mathrm{non-Ia}}$(H12)  } & SNN(Base)  & Fit (\citetalias{Hlozek_2012}) & 0.0048$\pm$0.0008 & -0.0015 (0.0338) \\
6) & \texttt{SNN(J17) D$_{\mathrm{non-Ia}}$(H12)  } & SNN(J17)  & Fit (\citetalias{Hlozek_2012}) & 0.0135$\pm$0.0012 &  0.0025 (0.0331) \\
7) & \texttt{SNN(DES-CC)   D$_{\mathrm{non-Ia}}$(H12)  } & SNN(DES-CC) & Fit (\citetalias{Hlozek_2012}) & 0.0048$\pm$0.0008 & 0.0070 (0.0329) \\
8) & \texttt{SNN(global) D$_{\mathrm{non-Ia}}$(Base) } & SNN(global)   & Baseline &  0.0128$\pm$0.0010 & 0.0253(0.0319) \\
9) & \texttt{SNN(randHost) D$_{\mathrm{non-Ia}}$(Base) } & SNN(randHost)  & Baseline & 0.0043$\pm$0.0007 & 0.0095 (0.0328) \\
\hline
10) & \texttt{\UnityClass} & P$_{\text{Ia}}$=1 $\forall$ SN & $\ddagger$ &  -0.0252$\pm$0.0046 & 0.0407 (0.0517) \\
11) & \texttt{\UnityClass+Chauvenet} & P$_{\text{Ia}}$=1 $\forall$ SN & $\ddagger$ & -0.0152$\pm$0.0014 & -0.0018 (0.0346) \\
12) & \texttt{\UnityClass+Chauvenet, $c$<0.15} & P$_{\text{Ia}}$=1 $\forall$ SN & $\ddagger$ & -0.0139$\pm$0.0020 & -0.0005 (0.0345) \\
\hline
\end{tabular}
    \begin{tablenotes}\footnotesize
    \item (a) The numbers of selected SNe are in Table~\ref{table:cuts}. The SALT2 selection and the requirement of a valid bias correction is always applied. Any additional selection criteria are indicated in the name of the BBC configuration.
    \item (b) Calculated using equation~\ref{eq:bias_definition}.
    \item (c) Biases measured from the DES-SN sample. Shifts are with respect to the value estimated using our BBC reference \texttt{SNN(Base) D$_{\mathrm{non-Ia}}$(Base)}. Errors reported in parenthesis are the \textit{statistical} uncertainties on $w$ only.
    \item $\ddagger$ Assuming all SNe have $P_{\mathrm{Ia}}=1$ means that the core collapse SN term in the BEAMS likelihood is always zero (equation~\ref{eq:likelihoods}). 
    \item ${*}$ Reference BBC configuration. For this BBC configuration, we obtain $\Delta w$ of 0.0045$\pm$0.0008 for Baseline simulation, 0.0082$\pm$0.0008 for LFs+Offset simulation,  0.0046$\pm$0.0009 for Dust(H98) simulation, 0.0019$\pm$0.0007 for J17 and 0.0076$\pm$0.0009 for DES-CC. 
    \end{tablenotes}
\end{table*}

\subsection{Biases for a flat $w$CDM model}
\label{sec:discussion}

\begin{figure*}
    \includegraphics[width=\linewidth]{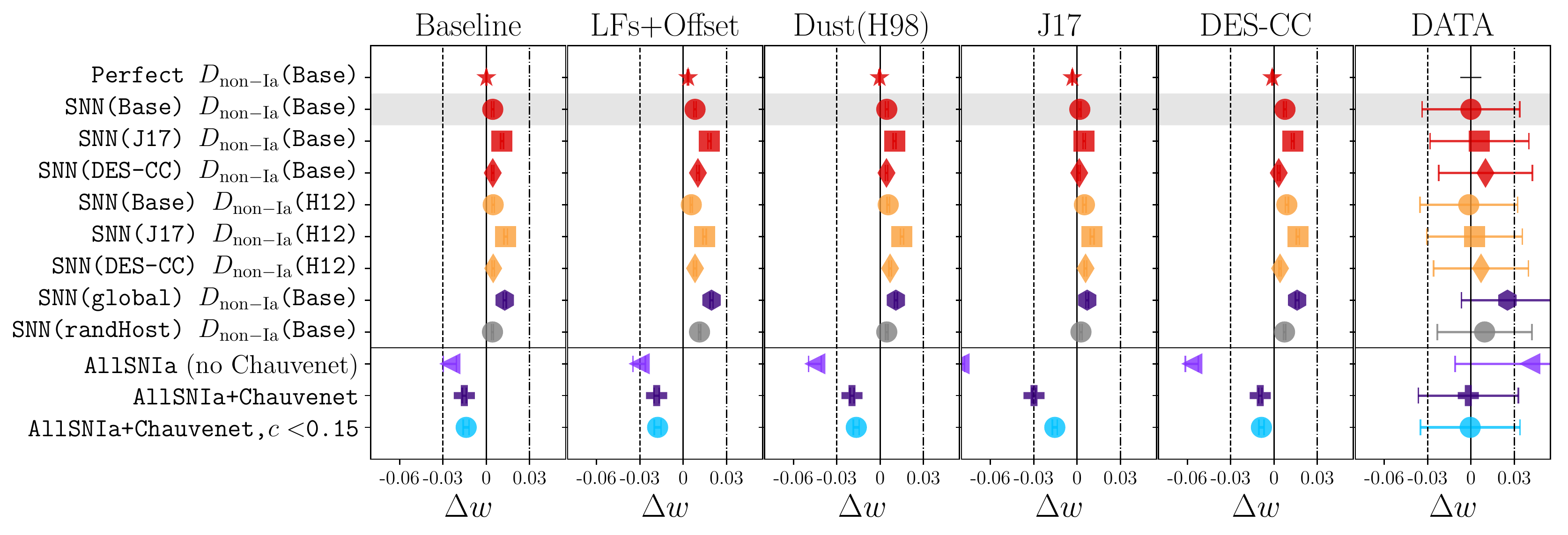}
\caption{Biases on the recovered dark energy equation-of-state parameter, $\Delta w$, measured for each simulation (Table~\ref{table:sims}) and for different BBC configurations (Table~\ref{table:BBC_opts}). Different \snn\ models correspond to different symbols (circles for \texttt{SNN(Base)}, squares for \texttt{SNN(J17)}, diamonds for \texttt{SNN(DES-CC)}), and different $D_{\mathrm{non-Ia}}$ modelling approaches correspond to different colours (red for \texttt{D$_{\mathrm{non-Ia}}$(Base)}
and orange for \texttt{D$_{\mathrm{non-Ia}}$(H12)}). For simulations, we estimate $\Delta w$ and relative uncertainties as described in equation~\ref{eq:bias_definition}. For the data (last column), we present $\Delta w$ with respect to our reference BBC configuration (\texttt{SNN(Base)\,$D_{\texttt{CC}}$(Base)}, second from the top, highlighted). Data error bars are 1$\sigma$ statistical uncertainties only, and are not independent for each BBC configuration.}
    \label{fig:wbias_SNN}
\end{figure*}

We first consider fits in a $w$CDM model.
Our key results are in Fig.~\ref{fig:wbias_SNN}, showing $\Delta w$ estimated using different BBC options and simulations. The cosmological results presented from the data are preliminary and are blinded (i.e., the best-fitting cosmology is not known) and are therefore also shown as shifts $\Delta w$ with respect to the (arbitrary) BBC reference configuration (\texttt{SNN(Base)\,D$_{\text{CC}}$(Base)}). Uncertainties on the data are the 1$\sigma$ statistical uncertainties, while for simulations we average the results of 50 realizations.

\begin{figure*}
   \centering
\includegraphics[width=0.7\linewidth]{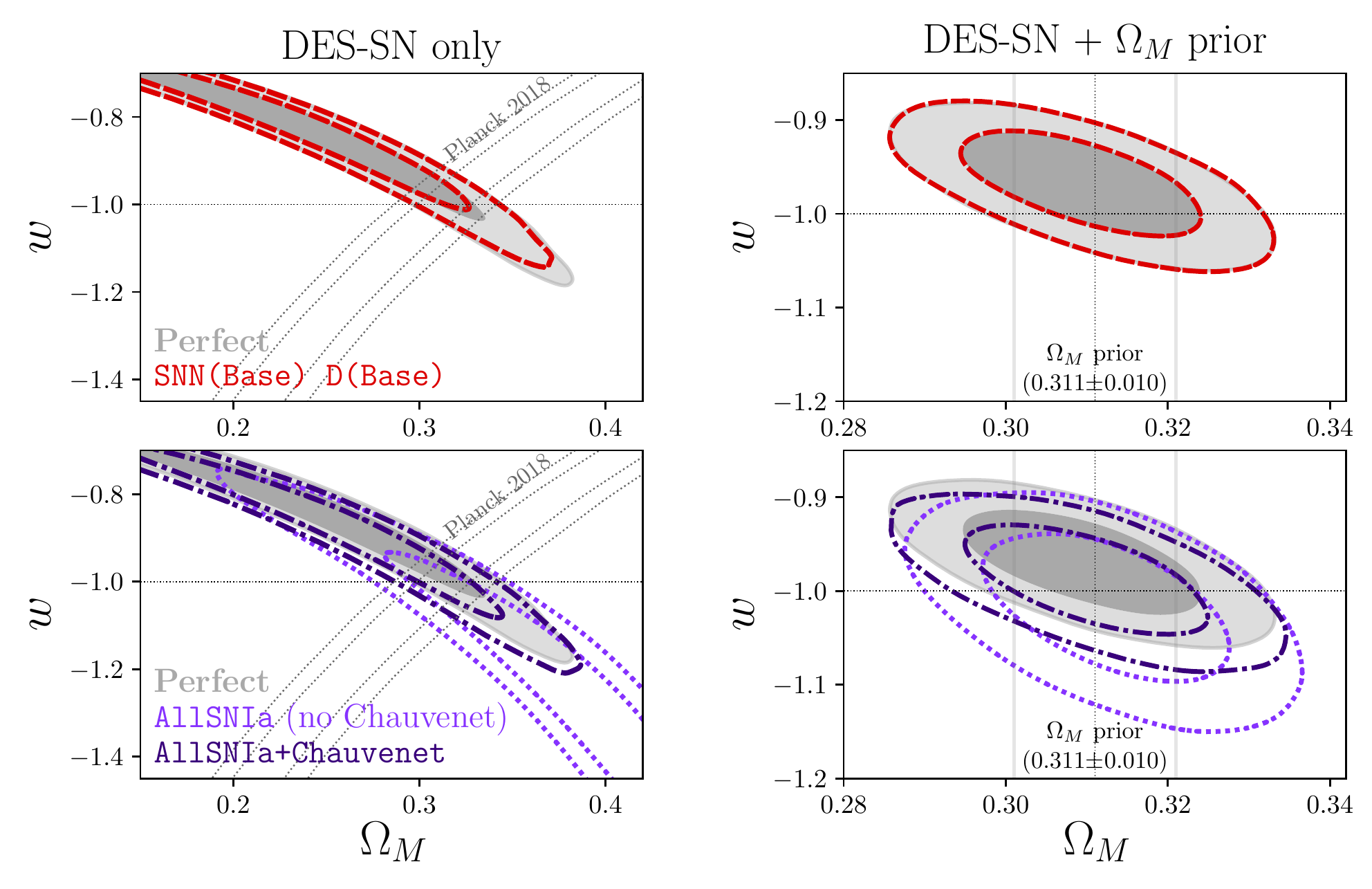}
    \caption{Cosmological contours estimated from a DES-SN simulated sample. The results show one realization of the Baseline simulation. We show the cosmological constraints (68 and 95 per cent confidence intervals) on $\Omega_M-w$ for a flat $w$CDM model, with and without an $\Omega_M$ prior of $0.311\pm0.010$ (right and left panels respectively). We also present results obtained using a perfectly classified sample of SNe Ia (grey filled contours), a contaminated sample of SNe analyzed assuming all SNe passing SALT2 selection are SNe Ia (dotted purple contours, lower panel), assuming all SNe passing SALT2 selection and Chauvenet's criterion are SNe Ia (dotted-dashed purple contours, lower panel) and using our reference BBC configuration \texttt{SNN(Base) D$_{\mathrm{non-Ia}}$(Base)} (dashed red contours, top panel). The $\Omega_M$ prior of $0.311\pm0.010$ is in grey.}
    \label{fig:contours}
\end{figure*}

\begin{figure*}
\centering
\includegraphics[width=0.92\linewidth,left]{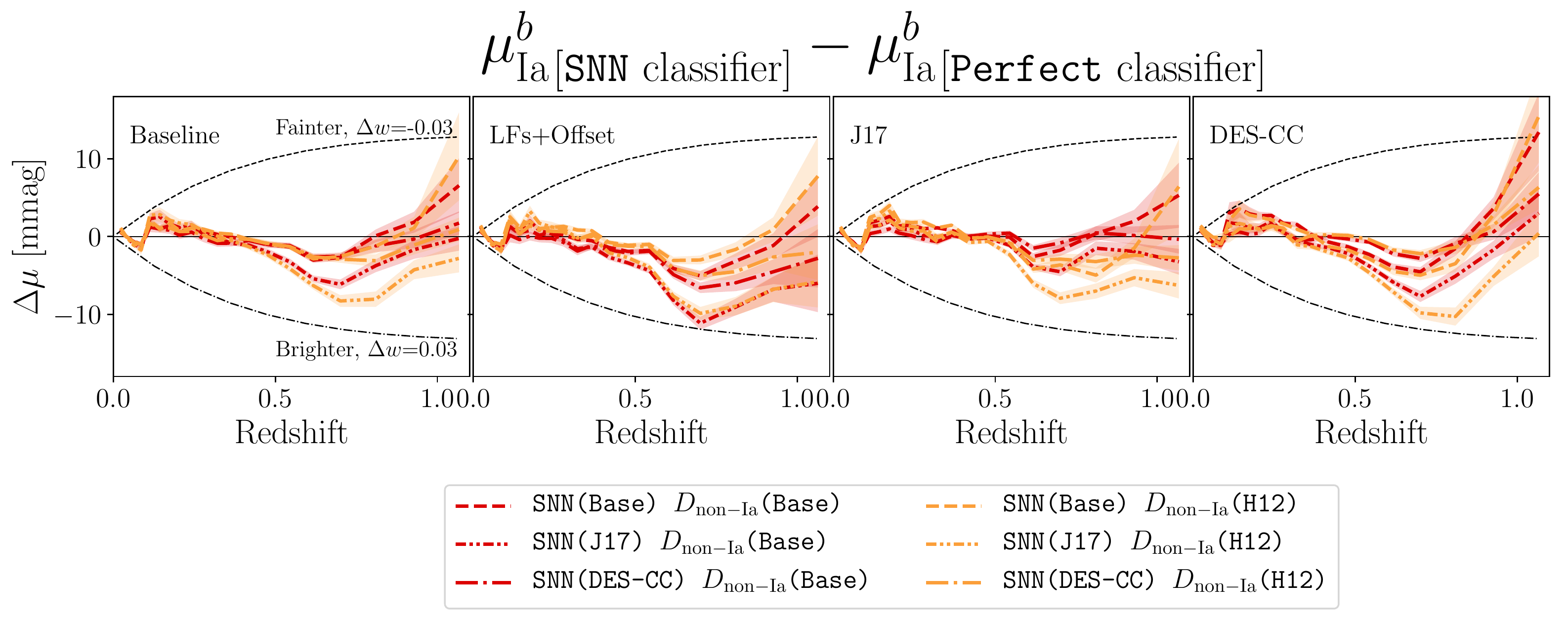}
\caption{Differences in the binned distance modulus $\mu_{\mathrm{Ia}}^{\BinInd}$ when using \snn\ compared to using the \texttt{Perfect} classifier. Each panel presents the results when applied to a different simulation: Baseline (left), LFs+Offset (centre left), J17 (centre right) and DES-CC (right). We compare different \snn\ classifiers and different BBC configurations: each \snn\ model corresponds to a different line-style, and each $D_{\mathrm{non-Ia}}$ modelling approach corresponds to a different colour (see legend).
Differences in distance modulus between $\Delta w=-0.03$ and $\Delta w=0.03$ are presented as dashed and dotted-dashed lines respectively.}
\label{fig:compare_mudif}
\end{figure*}

\begin{figure*}
\centering
\includegraphics[width=0.92\linewidth,left]{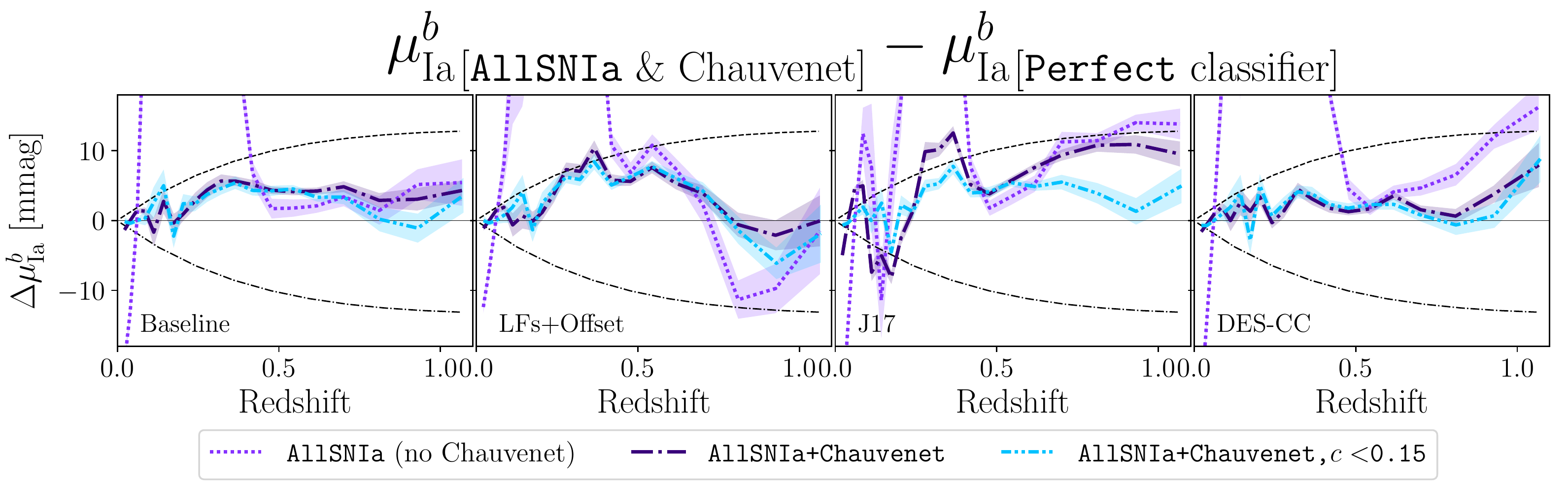}
\caption{As Fig.~\ref{fig:compare_mudif}, but comparing the \texttt{\UnityClass} approach with perfect classification. We combine the \texttt{\UnityClass} approach with different SN selection criteria: SALT2 selection only (\texttt{\UnityClass}), SALT2 selection and Chauvenet's criterion (\texttt{\UnityClass+Chauvenet}), and finally including stricter SALT2 $c$ cuts (\texttt{\UnityClass+Chauvenet,$c<0.15$}).}
   \label{fig:compare_mudif_chi2}
\end{figure*}

\begin{figure*}
\centering
\includegraphics[width=\linewidth,left]{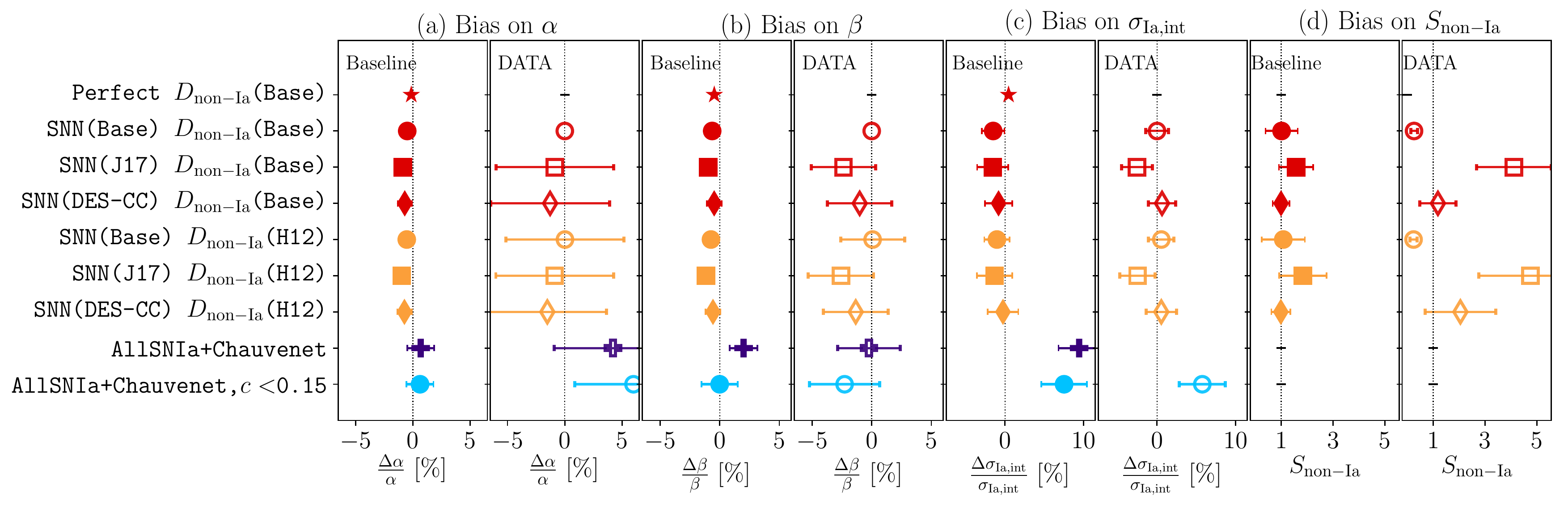}
\caption{Relative differences in the fitted nuisance parameters $\alpha$ (panel \textbf{(a)}), $\beta$ (panel \textbf{(b)}) and $\sigma_{\mathrm{Ia, int}}$ (panel \textbf{(c)}) and the deviation from one of the scaling parameter $S_{\mathrm{non-Ia}}$ (panel \textbf{(d)}). Each pair of panels presents the results for Baseline (left) and DES-SN5YR data (right). We compare different BBC configurations (see Table~\ref{table:BBC_opts}).  The BBC fitting procedure does not return uncertainty on $\sigma_{\mathrm{Ia, int}}$. Therefore, for both data and simulations the uncertainties on $\sigma_{\mathrm{Ia, int}}$ are estimated as the rms spread in $\sigma_{\mathrm{Ia, int}}$ measured from the 50 realizations of the Baseline simulation.}
   \label{fig:compare_alpha_beta}
\end{figure*}

\subsubsection{Cosmological biases using the \snn\ classifier}
\label{sec:biases_SNN}

Testing the different simulations presented in Section~\ref{sec:simulations} with \snn, we find that the biases on $w$ are $<1$ per cent (from a minimum of $\Delta w=0.002$ for Baseline simulation to a maximum $\Delta w=0.008$ for J17 simulation) for our BBC reference configuration, and $<2$ per cent for the other configurations in Table~\ref{table:BBC_opts} (a maximum $\Delta w=0.015$ is estimated for LFs+Offset simulation analyzed with \texttt{SNN(J17)} model). Across all the BBC configurations and simulations tested, the biases on the fitted nuisance parameters $\alpha$ and $\beta$ are $<1.5$ and $<1.8$ per cent respectively (see Fig.~\ref{fig:compare_alpha_beta}). Biases on SN Ia intrinsic scatter $\sigma_{\mathrm{Ia, int}}$ are also consistent with zero and the recovered scaling parameter $S_{\mathrm{non-Ia}}$ is consistent with one.

In Fig.~\ref{fig:contours}, we present the full $\Omega_M-w$ cosmological contours\footnote{As described in Sec.~\ref{sec:cosmo_estimation}, we estimate contours using the cosmological fitter CosmoMC.} from a single realization of the DES-like sample (i.e., the same statistical constraining power as expected from the DES-SN photometric sample). We compare cosmological contours for the ideal scenario of a perfectly classified sample of SNe Ia and for the realistic scenario of a contaminated sample of SNe Ia analysed using the \snn\ classifier. The biases on cosmological constraints due to contamination are significantly smaller than the statistical uncertainties.

Fig.~\ref{fig:compare_mudif} shows the biases on the binned Hubble diagram ($\Delta \mu$) using  different \snn\ models. Generally, the $|\Delta \mu|$ are less than 10\,mmag across all tests and simulations (consistent with the small biases measured on $w$). We observe consistently across all simulations that SN Ia distances estimated from BBC are mostly unbiased ($\Delta \mu<$4\,mmag) at lower redshifts ($z<0.5$), and the largest biases are observed at $z\sim0.7$, towards negative values (i.e., brighter values). At these redshifts, the number of true SNe Ia decreases and thus the modelling of the core collapse SN population is both more critical and more uncertain. This makes the marginalisation of core collapse SN contamination from BBC less accurate. The choice of the modelling approach adopted for the contamination likelihood can have a significant impact on $\mu_{\mathrm{Ia}}^{\BinInd}$.
For the same \snn\ model, $\mu_{\mathrm{Ia}}^{\BinInd}$ can differ by $>5$\,mmag when varying the modelling of the contamination likelihood. This is particularly evident in the simulation where contaminants are artificially brightened (LFs+Offset).  This suggests that the choice of training sample for \snn\ is not the only driver of systematics.

Finally, we note that for all our tests with \snn\, we find that the binned Hubble diagram $\mu_{\mathrm{Ia}}^{\BinInd}$ is mainly biased towards negative values, and this in turn corresponds to positive biases on $w$. This suggests that combining \snn\ with the BEAMS formalism tends to slightly \lq over-correct\rq\ for contamination and, therefore, preferentially biases the Hubble diagram towards brighter values. In the next section, we discuss cosmological biases when applying Chauvenet's criterion and no classification and we observe the opposite trend.

\subsubsection{Cosmological biases using Chauvenet's criterion without a classifier}
\label{sec:biases_Chauvenet}

We next test the case of not using a classifier and assuming all SNe in the samples that pass the SALT2 selection are SNe Ia (\texttt{\UnityClass}), setting $P_{\mathrm{Ia}}=1$ for every SN and the contamination term in the BEAMS likelihood to zero. We also test outlier rejection in combination with the \texttt{\UnityClass} approach, with the results in Fig.~\ref{fig:compare_mudif_chi2}.

With no outlier rejection, the binned $\mu_{\mathrm{Ia}}^{\BinInd}$ are biased towards fainter values pulled by faint core collapse SN contaminants, especially at $z<0.5$. At higher-$z$ the biases are smaller ($<10$\,mmag) as contamination is naturally reduced by Malmquist bias, and can either be brighter (e.g., for LFs+Offset) or fainter (e.g., J17) depending on the properties of the simulated core collapse SNe. As expected, this approach results in significant biases with $\Delta w = -0.025\pm0.009$ for Baseline up to $\Delta w=-0.082\pm0.008$ for J17 (see also Fig.~\ref{fig:contours}). 
The biases from this no-classifier approach have the opposite sign compared to the biases found when combining \snn\ and the BEAMS approach. In the no-classifier approach, the fainter population of contamination is \lq under-corrected\rq\ (or effectively not corrected at all as core collapse SNe are assume to have $P_{\mathrm{Ia}}=1$), therefore the biases on $\mu_{\mathrm{Ia}}^{\BinInd}$ are mainly positive and $w$-bias is negative.

When we combine Chauvenet's criterion with \texttt{\UnityClass}, the biases in $\mu_{\mathrm{Ia}}^{\BinInd}$ are reduced, generally to $<10$\,mmag, and are broadly consistent with the \snn\ results (Fig.~\ref{fig:compare_mudif_chi2}). The $w$-biases range from $-0.010\pm0.002$ for Baseline to $-0.019\pm0.001$ for Dust(H98) 
(Fig.~\ref{fig:wbias_SNN}). However, in the J17 simulations, while the fraction of contaminants (mostly red type Ib SNe) is similar to the other simulations (Table~\ref{tab:contam_table_allSNIa_postBBC}), their distribution on the Hubble diagram is such that, even after applying Chauvenet's criterion, a significant trend in $\mu_{\mathrm{Ia}}^{\BinInd}$ is introduced biasing $w$ by $-0.030\pm0.004$. This is reduced by 50 per cent with a stricter SALT2 $c$ selection (to $-0.015\pm0.02$), suggesting that the bulk population of red and bright contaminants is the main driver of this cosmological bias. For the other simulations, applying stricter SALT2 $c$ cuts does not reduce biases on $w$ significantly, while it reduces the number of SNe Ia by 8 per cent. 

Fig.~\ref{fig:compare_alpha_beta} shows that the fitted nuisance parameters are also biased when using Chauvenet's criterion only. When applying Chauvenet's criterion, the residual population of red and faint core-collapse contaminants lead to an overestimate of the fitted values of $\beta$ by approximately 3 per cent. These biases are reduced to $<1$ per cent when applying stricter SALT2 $c$ cuts. Biases on $\alpha$ are $<1$ per cent. The SN Ia intrinsic scatter is also overestimated by 7 to 10 per cent.

The cosmological constraints presented in Fig.~\ref{fig:contours} highlight the power of outlier rejection methods like Chauvenet's criterion. For a DES-like simulated sample, when we assume all SNe passing SALT2 selection and Chauvenet's criterion are SNe Ia (\texttt{AllSNIa+Chauvenet}), the biases on the cosmological contours are small.
These findings and the results presented Fig.~\ref{fig:wbias_SNN} and Fig.~\ref{fig:compare_mudif_chi2} suggest that cosmological biases due to contamination can be small even without applying photometric classification algorithms and using only outlier rejection methods.

\subsubsection{The role of priors}
\label{sec:priors}
Besides \snn\ and Chauvenet's criterion, the $\Omega_M$ prior discussed in Section~\ref{sec:cosmo_estimation} is another element that indirectly contributes to reduce biases on $w$ due to contamination. In SN cosmology, SNe Ia measurements and CMB measurements are typically combined in order to break the respective degeneracies on the $\Omega_M$ and $w$ parameter space, and thus reduce the overall statistical uncertainty on $w$. As shown in Fig.~\ref{fig:contours} (left panel), core collapse contamination shifts the SN-only cosmological contours \emph{along} the \lq banana-shaped\rq\ SN contours and \emph{perpendicularly} to the CMB constraints and to a Gaussian $\Omega_M$ prior. Therefore, combining SNe with CMB measurements (left panels in Fig.~\ref{fig:contours}) or applying an $\Omega_M$ prior (right panels in Fig.~\ref{fig:contours}) not only reduces statistical uncertainties on $w$, but also significantly mitigates systematic biases on $w$ due to contamination.

We highlight that, for $w$ estimates, CMB constraints are more stringent (i.e., almost perfectly orthogonal to SN-only constraints) than a Gaussian $\Omega_M$ prior. For this reason, we anticipate that updating our prior with the latest CMB measurements from \citet{collaboration2018planck} will further reduce statistical uncertainties on $w$ and systematic biases on $w$ due to contamination.

\subsubsection{Biases when applied to data}
\label{sec:biases_data}

We perform the same tests on the DES-SN data as applied to the simulations. Clearly, the true classification of each SN and the unbiased $\mu_{\mathrm{Ia}}^{\BinInd}$ is not known, so we estimate relative biases between different BBC configurations.

Table~\ref{table:BBC_opts} (last column on the right) and Fig.~\ref{fig:wbias_SNN} (last column on the right) present $\Delta w$ shifts measured from the data and estimated with respect to the value of $w$ fitted from our reference BBC configuration. Using Chauvenet's criterion and assuming all events are SNe Ia, we obtain $\Delta w=-0.0018$ (r.m.s. on $\Delta w$ estimated from 50 realizations of the Baseline simulation is 0.0076). This result suggests that our reference BBC configuration and the Chuavenet's criterion approach are consistent within the uncertainties. When comparing our reference BBC configurations with the BBC configurations that use \snn\ models \texttt{SNN(J17)} and \texttt{SNN(DES-CC)} (i.e., BBC configurations 3 and 4 in Table~\ref{table:BBC_opts}), we observe shifts on $w$ of 0.0059 (r.m.s. from simulations is 0.0036) and 0.0101 (r.m.s. from simulations is 0.0036). The BBC configuration that implements the \texttt{SNN(global)} classifier results in the largest $\Delta w$, but given the caveats discussed in Section~\ref{sec:cont_eff_globalcosmo}) we do not consider \texttt{SNN(global)} a robust classification method. The statistical uncertainty on $w$ for our reference BBC configuration is 0.034, which is approximately three times the maximum $\Delta\mu$ observed in the data. These results confirm that for the cosmological analysis of the DES photometric SN sample, contamination is a subdominant systematic when compared to the statistical uncertainty.

In Fig.~\ref{fig:compare_alpha_beta}, we compare fitted nuisance parameters when using the reference BBC configuration and other BBC configurations. The parameters $\alpha$ and $\beta$ fitted from the data are consistent between the different configurations tested. Large discrepancies are seen in the fitted values of the scaling factor $S_{\mathrm{non-Ia}}$. $S_{\mathrm{non-Ia}}$ for the data is 0.26$\pm$0.13, 4.11$\pm$1.44 and 1.18$\pm$0.70 when using \texttt{SNN(Base)}, \texttt{SNN(J17)} and \texttt{SNN(DES-CC)} respectively and the non-Ia likelihood approach \texttt{$D_{\mathrm{non-Ia}}$(Base)}. Predicting \texttt{$D_{\mathrm{non-Ia}}$} and constraining the factor $S_{\mathrm{non-Ia}}$ is difficult when the percentage of contaminants in the sample is already very low and this explains these large differences in the fitted values.

For comparison and a sanity check, we also test the performances of the \snn\ classifier \texttt{SNN(Base)} and Chauvenet's criterion on the DES-SN sample of spectroscopically-confirmed SNe. After applying all the selection criteria discussed in Section~\ref{sec:salt2_cut}, we have 401 spectroscopically-classified SNe observed by DES. We find that 354 events are certain SNe Ia, 44 likely SNe Ia and 3 are classified as non-Ia (two stripped envelope SNe and one hydrogen-rich SN). Only one out of the three non-Ia SNe satisfy Chauvenet's criterion. All three events have $P_{\text{Ia}}<0.2$. The spectroscopic sample is significantly biased towards bright, high signal to noise ratio events, therefore it is not surprising that the contamination is extremely low (less than 1 per cent after SALT2-based cuts only and zero after probability cuts). However, it shows how efficiently a SALT2-based selection and Chauvenet's criterion can reduce contamination, as generally applied in the cosmological analysis of spectroscopic samples of SNe Ia \citep{scolnic2018, DES_syst, Foley_Foundation}.

\begin{figure}
\centering
\includegraphics[width=0.82\linewidth]{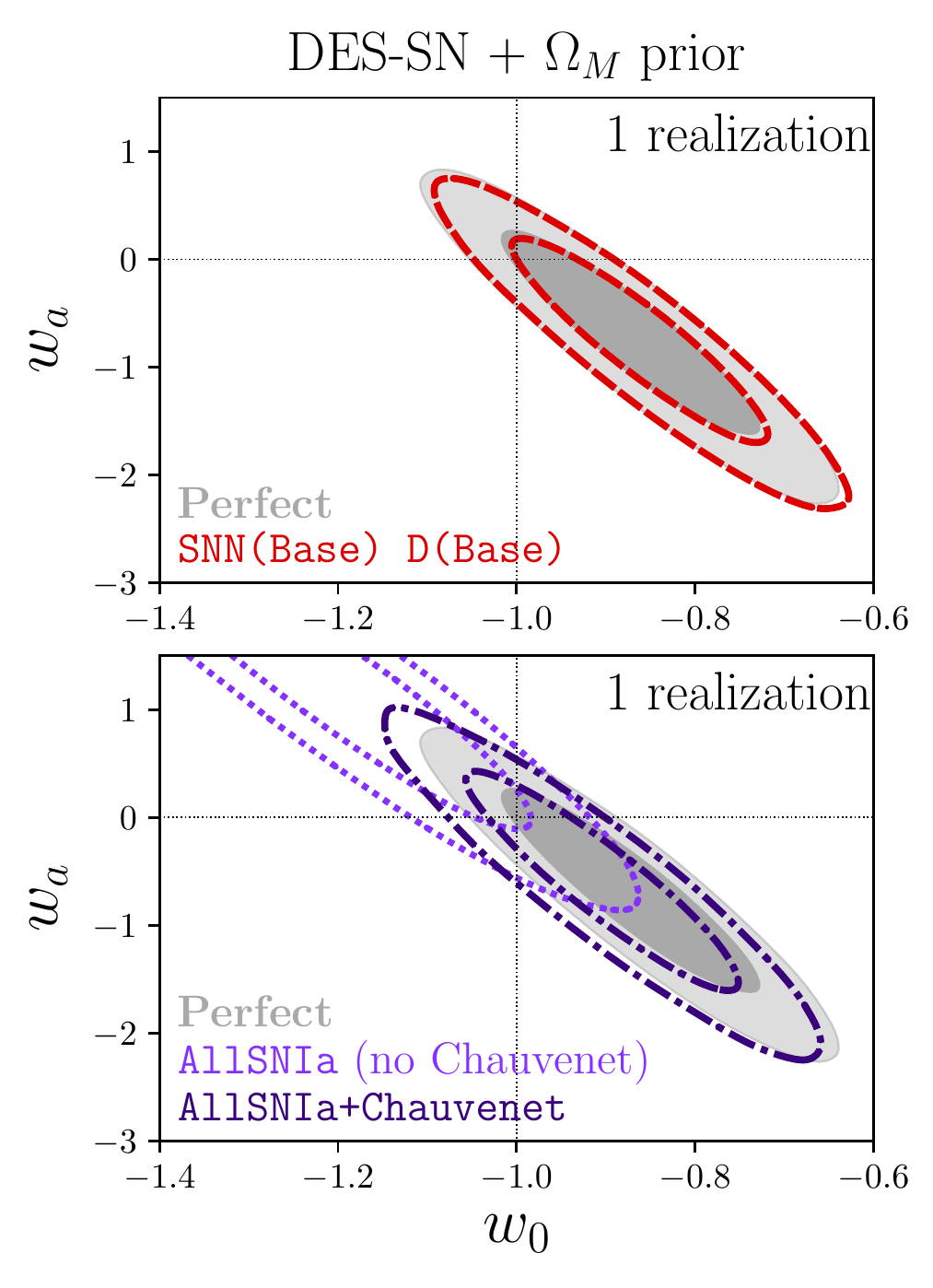}\\
    \caption{Same as Fig.~\ref{fig:contours} but in the $w_0-w_a$ plane and assuming a flat $w_0w_a$CDM model with an $\Omega_M$ prior of $0.311\pm0.010$.}
    \label{fig:contours_w0wa}
\end{figure}

\subsection{Systematic uncertainties associated with contamination}

In this section, we estimate the contribution of contamination to the $w$ systematic error budget from a DES-like cosmological analysis. In order to do this, we follow the approach presented by \citet{2011ApJS..192....1C} and \citet[][section 3.8.2]{DES_syst} and define a systematic covariance matrix, $C_{\mathrm{syst}}$, that can be included in the fit for cosmological parameters. The $\chi^2$-minimization cosmological fitter introduced in Sec.~\ref{sec:cosmo_estimation} does not currently handle a systematic covariance matrix; for this reason, we use CosmoMC when estimating systematic uncertainties on $w$.

Given $\partial \mu_{\mathrm{Ia, s_k}}^{\BinInd}$ the differences in the binned Hubble diagram after changing the systematic parameter $s_k$, the systematic covariance matrix, $C_{\mathrm{syst}}^{ij}$, is defined as
\begin{equation}
    C_{\mathrm{syst}}^{ij} = \sum_{k=1}^{N_{\mathrm{syst}}} \left( \frac{\partial \mu_{\mathrm{Ia, s_k}}^i}{\partial s_k} \right) \left( \frac{\partial \mu_{\mathrm{Ia,s_k}}^j}{\partial s_k} \right) \sigma^2_{s_k},
    \label{eq:cov_syst}
\end{equation}
where $\sigma_{s_k}$ is the uncertainty of the systematic $s_k$ and the indexes $i$ and $j$ are iterated over the $N_{\mathrm{bins}}$ redshift bins ($i,j=1,...,N_{\mathrm{bins}}$). 

We build two different covariance matrices: one that includes variations over the three \snn\ models (\texttt{SNN(Base)}, \texttt{SNN(J17)} and \texttt{SNN(DES-CC)}) but fixes the contamination likelihood to \texttt{$D_{\mathrm{non-Ia}}$(Base)} (configurations 2, 3 and 4 in Table~\ref{table:BBC_opts}), and one that includes variations over the three \snn\ models (\texttt{SNN(Base)}, \texttt{SNN(J17)} and \texttt{SNN(DES-CC)}) but fixes the contamination likelihood to \texttt{$D_{\mathrm{non-Ia}}$(H12)} (configurations 5, 6 and 7 in Table~\ref{table:BBC_opts}).
For each systematic, we estimate the contribution to the total error budget on $w$ by applying the definition presented by \citet[][equation 22]{DES_syst}
\begin{equation}
    \sigma'_w = \sqrt{(\sigma_{\mathrm{stat+syst}}^2 - \sigma_{\mathrm{stat}}^2)},
\end{equation}
where $\sigma_{\mathrm{stat+syst}}$ is the uncertainty estimated when considering only one (or a sub-group of) systematics and $\sigma_{\mathrm{stat}}$ is the statistical uncertainty. The results are estimated for our Baseline simulation and presented in Table~\ref{table:syst_budget} (and obtain similar results when performing the same test on the other simulations).
Systematic uncertainties associated with contamination are 0.004 for the \texttt{$D_{\mathrm{non-Ia}}$(Base)} method and 0.007 for the polynomial fitting method by \citetalias{Hlozek_2012}. In general, systematics associated with contamination are at most a third of the statistical error, which corresponds to an increase of the overall $w$ error budget by less than 5 per cent. 

In Appendix~\ref{appendix_P05}, we highlight some potential limitations related to the \texttt{$D_{\mathrm{non-Ia}}$(H12)} approach and to the choice of modelling the core-collapse likelihood term as a second order polynomial. Therefore, we consider the \texttt{$D_{\mathrm{non-Ia}}$(Base)} method as the most reliable one in our analysis and quote $\sigma_w'=0.004$ to be our best estimate of systematic uncertainties associated with contamination. 

\begin{table}
\caption{Uncertainty contributions to $w$ for a $w$CDM model (SNe are combined with a $\Omega_M$ prior of $0.311\pm$0.010). See Table~\ref{table:BBC_opts} for a detailed description of the BBC configurations listed in the first column.}
\footnotesize
\label{table:syst_budget}
\begin{tabular}{lcccc}
\hline
 & $\sigma'_w$ & $\sigma'_w/\sigma_{\mathrm{stat}}$ & $\sigma_{\mathrm{stat+syst}}$ \\
\hline
Total $\sigma_{\mathrm{stat}}$ & - & - & 0.039 \\
\hline
2) \texttt{SNN(Base) D$_{\mathrm{non-Ia}}$(Base)} & & & \\
3) \texttt{SNN(J17) D$_{\mathrm{non-Ia}}$(Base)} & 0.004 & 0.106 & 0.040 \\
4) \texttt{SNN(DES-CC) D$_{\mathrm{non-Ia}}$(Base)} & & & \\
\hline
5) \texttt{SNN(Base) D$_{\mathrm{non-Ia}}$(H12)} & & & \\
6) \texttt{SNN(J17) D$_{\mathrm{non-Ia}}$(H12)} & 0.007 & 0.171 & 0.040 \\
7) \texttt{SNN(DES-CC) D$_{\mathrm{non-Ia}}$(H12)} & & & \\
\hline
\end{tabular}
\end{table}

\subsection{Biases for a time-varying $w_0$/$w_a$ model}
\begin{figure*}
\includegraphics[width=\linewidth]{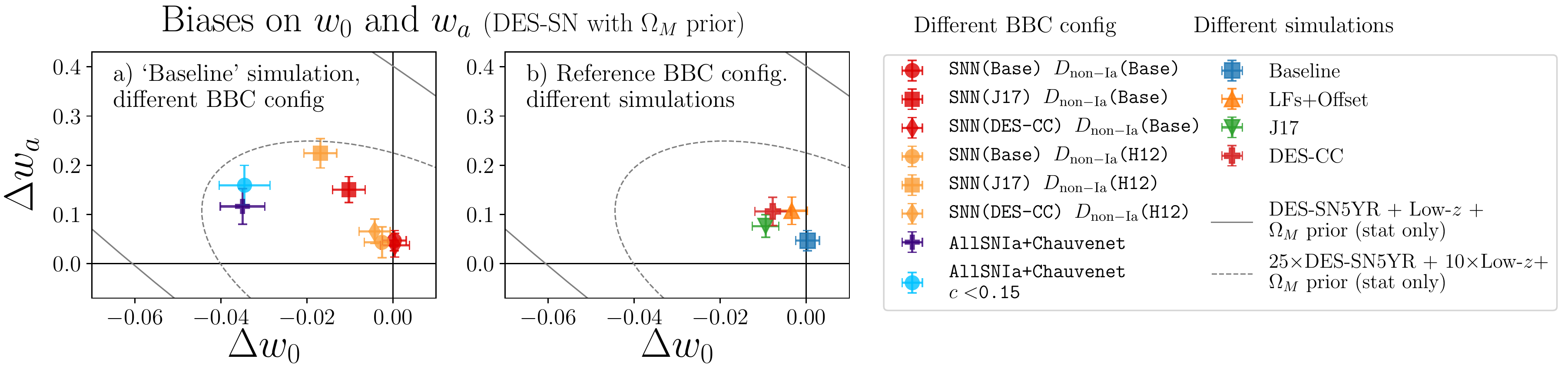}\\
\caption{Biases on the dark energy equation-of-state parameters $w_0$ and $w_a$ measured using \textit{(a)} the Baseline simulation and varying the BBC configuration, and \textit{(b)} the reference BBC configuration (Table~\ref{table:BBC_opts}) but varying the core collapse SN simulation (panel \textbf{(b)}). In panel \textit{(a)}, different \snn\ models correspond to different symbols (circles for \texttt{SNN(Base)}, squares for \texttt{SNN(J17)}, diamonds for \texttt{SNN(DES-CC)}), and different $D_{\mathrm{non-Ia}}$ modelling approaches correspond to different colours (red for \texttt{D$_{\mathrm{non-Ia}}$(Base)} and orange for \texttt{D$_{\mathrm{non-Ia}}$(H12)}). In panel \textit{(b)}, we present results for four different core collapse simulations (see Table~\ref{table:sims}). The biases $\Delta w_0$ and $\Delta w_a$ and the relative uncertainties are measured as described in equation~\ref{eq:bias_definition}. As a comparison, we show the zero-centred 68\% $w_0-w_a$ contours from DES-SN sample combined with an $\Omega_M$ prior of $0.311\pm0.010$ and from a sample of 25 times the size of DES-SN sample. }
    \label{fig:w0wabias}
\end{figure*}

We analyze the effects of contamination when fitting our simulated SN samples assuming a flat $w_0w_a$CDM model. 
In Fig~\ref{fig:contours_w0wa}, we present the $w_0-w_a$ cosmological contours obtained from one realization of the Baseline simulation and assuming a Gaussian $\Omega_M$ prior of $0.311\pm0.010$. 

In Fig.~\ref{fig:w0wabias}, we present the average biases on $w_0$ and $w_a$ measured for the Baseline simulation. For different BBC configurations (Table~\ref{table:BBC_opts}) and \snn, we find a $-0.011$ to 0.001 bias on $w_0$ and 0.008 to 0.166 bias on $w_a$. Using Chauvenet's criterion and \texttt{AllSNIa}, we find biases of $-0.031$ and 0.097 on $w_0$ and $w_a$ respectively. If we assume our reference BBC configuration is the most robust one, we measure biases across the different core collapse SN simulations of $-0.009<\Delta w_0<0.000$ and $0.047<\Delta w_a<0.108$. This is shown in Fig.~\ref{fig:w0wabias}. 

By comparison, the average statistical uncertainties on $w_0$ and $w_a$ expected for a DES-like sample are $0.097$ and $0.620$, i.e., 5 to 10 times larger than the biases $\Delta w_0$ and $\Delta w_a$ due to contamination. 

Looking further to the future, these results can inform the planning of future time-domain experiments such as the optical Legacy Survey of Space and Time \citep[LSST;][]{2019ApJ...873..111I} that will be conducted using the Vera Rubin Observatory. Although the exact observational strategy is being developed, LSST is expected to discover more than 1000 new SNe Ia per night. Spectroscopic follow-up programmes such as the Time-Domain Extragalactic Survey \citep[TiDES;][]{2019Msngr.175...58S} and others, will provide host galaxy spectroscopic redshifts as well as spectroscopic classifications for a subset of these events. The photometric SN Ia sample is expected to include at least 25 times more cosmologically-useful SNe Ia than the DES-SN photometric SN Ia sample, with similar redshift distributions (Frohmaier et al. in prep.). In parallel, low redshift SN samples are also expected to increase  \citep[approximately $\times10$ more SNe Ia than available in current low-$z$ samples; see DESC Science Requirements Document;][]{2018arXiv180901669T}.

Following these forecasts, we estimate the statistical uncertainties on $w_0$ and $w_a$ expected when combining 25$\times$ the DES-SN5YR photometric SN sample, 10$\times$ the current low-$z$ samples, and an $\Omega_M$ prior of 0.311$\pm$0.010.
These are found to be 0.03 and 0.19 for $w_0$ and $w_a$ respectively, i.e., approximately 3 and 2 times larger than the biases $\Delta w_0$ and $\Delta w_a$ found when applying our reference BBC configuration on the full range of simulations. The contours are presented in Fig. \ref{fig:w0wabias}. We conclude that contamination is not expected to degrade the figure of merit of the LSST SN Ia sample significantly, especially when implementing classification techniques like \snn.

\section{Conclusions}
\label{sec:conclusions}

In this paper, we have exploited state-of-the-art simulations of SN candidates detected by the Dark Energy Survey (DES) to quantify systematic effects in cosmological analyses introduced by the use of photometric SN classification methods. We focused on the testing of SuperNNova (\snn), a SN photometric classification tool based on machine learning techniques. In order to provide a robust assessment of the algorithm's performance and avoiding potential over-fitting, we have trained and tested \snn\ not only on our \lq Baseline\rq\ simulation of DES (Table~\ref{table:sims}), but on a wider suite of DES simulations designed to explore different astrophysical assumptions in the core collapse SN population and different compilations of core collapse SN templates. We then perform a state-of-the-art analysis using SALT2 light curve fitting, BEAMS and its extension BBC to estimate bias corrections and correct for contamination, and cosmology fitting. In this way, we can propagate the effects of contamination to cosmological parameter estimation.
Our main findings are: 
\begin{itemize}
    \item Across our DES simulations, contamination ranges from 0.8--3.2 per cent when using \snn, with the efficiency of the classification ranging from 99.0--99.5 per cent (Table~\ref{tab:contam_table_allSNIa}). Therefore, on a sample of approximately 1680 SNe (Table~\ref{table:cuts}), we expect \snn\ to misclassify as SN Ia approximately 14 to 55 core collapse SNe and to exclude from the cosmological fit 9 to 17 true SNe Ia.
    \item \snn\ trained on our Baseline simulation performs well across all simulated data samples, including those based on independent libraries of core collapse SN templates, with a contamination of $\leq1.4$ per cent. \snn\ classifiers trained on simulations using templates from \citetalias{Jones_2017_I} or \citetalias{DES-CC} perform well when tested on simulations built using the same set of templates ($<1$ per cent contamination), but when tested on simulations built using independent core-collapse SN templates, contamination increases to $1.7-3.2$ per cent.

    \item Outlier rejection methods like Chauvenet's criterion can also significantly reduce contamination (to $<3.1$ per cent in the Baseline simulation, and $<5.3$ per cent for the other simulations, see Table~\ref{tab:contam_table}). This can be further reduced with a tighter selection based on the SN Ia colour ($<4.0$ per cent).

    \item We combine the BBC formalism with \snn\ trained on the Baseline simulation, and set this as our reference approach. Assuming a flat $w$CDM model, we find that biases on $w$ are below 1 per cent ($|\Delta w| <0.0082$). When exploring additional BBC configurations and \snn\ training methods, we find that biases on $w$ are at most $0.018$. These biases are respectively 4 and 2 times smaller than the expected statistical uncertainty on $w$ from DES-SN. The predicted systematic uncertainties related to contamination are $<0.007$ and this suggests that contamination increases by less than 5 per cent the total uncertainty on $w$ and it is not a limiting systematic for the cosmological analysis of the DES-SN sample. 

    \item When we implement Chauvenet's criterion and assume that all SNe that are not identified as outliers in the Hubble diagram are type Ia, this simplistic approach also provides relatively small biases on $w$ ($|\Delta w|<0.018$ and $|\Delta w|<0.033$ with and without stricter SALT2$c$-based selection). Thus, cosmological biases from contamination are small even without applying a photometric classification algorithm. 
    
    \item Core-collapse contamination shifts the SN-only cosmological contours perpendicularly to CMB constraints (see Fig.~\ref{fig:contours}). Therefore, combining SNe with CMB measurements (and not only with a Gaussian $\Omega_M$ prior) will not only reduce the statistical uncertainty on $w$, but also further mitigate systematic biases on $w$ due to contamination. In future cosmological analyses of the DES photometric SN sample, SN constraints will be combined with CMB constraints from \citet{collaboration2018planck}, therefore we anticipate our estimates of $w$-biases due to contamination and $\sigma_{\mathrm{stat}}$ on $w$ to decrease compared to using the Gaussian $\Omega_M$ prior in this paper. From a preliminary analysis, we forecast the contribution of contamination to the statistical error budget on $w$ (i.e., $\sigma'_w/\sigma_{\mathrm{stat}}$, see Table~\ref{table:syst_budget}) to change by less than 20 per cent.

    \item We estimate biases due to contamination on $w_0$ and $w_a$. Combing the DES-SN sample with a Gaussian $\Omega_M$ prior of $0.311\pm0.010$, we show the biases on $w_0$ to be less than 0.009, and the bias on $w_a$ to be less than 0.108. These are 5 to 10 times smaller than the statistical uncertainties on $w_0$ and $w_a$ expected from the DES-SN sample.
\end{itemize}


In general, the results in this paper are encouraging for the ongoing DES-SN cosmological analysis, and demonstrate the tools to fully exploit the photometric DES-SN sample to constrain the dark energy equation-of-state. Our work lays the foundation for the cosmological analysis of the DES photometric SN sample and our results will be essential to assess the systematic error budget on cosmological parameters estimated from the DES-SN sample. 


\section*{Acknowledgements}

This work was supported by the Science and Technology Facilities Council [grant number ST/P006760/1] through the DISCnet Centre for Doctoral Training. MS acknowledges support from EU/FP7-ERC grant 615929, and PW acknowledges support from STFC grant ST/R000506/1. TMD acknowledges support from ARC grant FL180100168. LG acknowledges financial support from the Spanish Ministry of Science, Innovation and Universities (MICIU) under the 2019 Ram\'on y Cajal program RYC2019-027683 and from the Spanish MICIU project PID2020-115253GA-I00.
RH and MS were supported by DOE grant DE-FOA-0001781 and NASA grant NNH15ZDA001N-WFIRST. The material is based upon work supported by NASA under award number 80GSFC17M0002. LK thanks the UKRI Future Leaders Fellowship for support through the grant MR/T01881X/1.

This paper has gone through internal review by the DES collaboration.
Funding for the DES Projects has been provided by the U.S. Department of Energy, the U.S. National Science Foundation, the Ministry of Science and Education of Spain, the Science and Technology Facilities Council of the United Kingdom, the Higher Education Funding Council for England, the National Center for Supercomputing Applications at the University of Illinois at Urbana-Champaign, the Kavli Institute of Cosmological Physics at the University of Chicago, the Center for Cosmology and Astro-Particle Physics at the Ohio State University, the Mitchell Institute for Fundamental Physics and Astronomy at Texas A\&M University, Financiadora de Estudos e Projetos, Funda{\c c}{\~a}o Carlos Chagas Filho de Amparo {\`a} Pesquisa do Estado do Rio de Janeiro, Conselho Nacional de Desenvolvimento Cient{\'i}fico e Tecnol{\'o}gico and 
the Minist{\'e}rio da Ci{\^e}ncia, Tecnologia e Inova{\c c}{\~a}o, the Deutsche Forschungsgemeinschaft and the Collaborating Institutions in the Dark Energy Survey. 

The Collaborating Institutions are Argonne National Laboratory, the University of California at Santa Cruz, the University of Cambridge, Centro de Investigaciones Energ{\'e}ticas, 
Medioambientales y Tecnol{\'o}gicas-Madrid, the University of Chicago, University College London, the DES-Brazil Consortium, the University of Edinburgh, 
the Eidgen{\"o}ssische Technische Hochschule (ETH) Z{\"u}rich, 
Fermi National Accelerator Laboratory, the University of Illinois at Urbana-Champaign, the Institut de Ci{\`e}ncies de l'Espai (IEEC/CSIC), 
the Institut de F{\'i}sica d'Altes Energies, Lawrence Berkeley National Laboratory, the Ludwig-Maximilians Universit{\"a}t M{\"u}nchen and the associated Excellence Cluster Universe, 
the University of Michigan, NFS's NOIRLab, the University of Nottingham, The Ohio State University, the University of Pennsylvania, the University of Portsmouth, 
SLAC National Accelerator Laboratory, Stanford University, the University of Sussex, Texas A\&M University, and the OzDES Membership Consortium.

Based in part on observations at Cerro Tololo Inter-American Observatory at NSF's NOIRLab (NOIRLab Prop. ID 2012B-0001; PI: J. Frieman), which is managed by the Association of Universities for Research in Astronomy (AURA) under a cooperative agreement with the National Science Foundation.

The DES data management system is supported by the National Science Foundation under Grant Numbers AST-1138766 and AST-1536171.
The DES participants from Spanish institutions are partially supported by MICINN under grants ESP2017-89838, PGC2018-094773, PGC2018-102021, SEV-2016-0588, SEV-2016-0597, and MDM-2015-0509, some of which include ERDF funds from the European Union. IFAE is partially funded by the CERCA program of the Generalitat de Catalunya.
Research leading to these results has received funding from the European Research
Council under the European Union's Seventh Framework Program (FP7/2007-2013) including ERC grant agreements 240672, 291329, and 306478.
We  acknowledge support from the Brazilian Instituto Nacional de Ci\^encia
e Tecnologia (INCT) do e-Universo (CNPq grant 465376/2014-2).

This manuscript has been authored by Fermi Research Alliance, LLC under Contract No. DE-AC02-07CH11359 with the U.S. Department of Energy, Office of Science, Office of High Energy Physics.

This work was completed in part with resources provided by the University of Chicago’s Research Computing Center.

Finally, this work was based in part on data acquired at the Anglo-Australian Telescope, under program A/2013B/012. We acknowledge the traditional owners of the land on which the AAT stands, the Gamilaraay people, and pay our respects to elders past and present.
\\
\newline
\textit{Software:}
\texttt{numpy} \citep{numpy}, \texttt{matplotlib} \citep{matplotlib}, \texttt{pandas} \citep{mckinney-proc-scipy-2010}, \texttt{scipy} \citep{2020SciPy-NMeth}, SNANA \citep{Kessler_2009}, Pippin \citep{Hinton2020}.
\newline 

\section*{Data availability statement}
Input and configuration files needed to train and test the photometric classifier \snn\ and to run BBC are available at \url{https://github.com/maria-vincenzi/DES_CC_simulations}. Data relative to the DES photometric sample used in Fig. \ref{fig:comparison} are also available.

\bibliographystyle{mnras}
\bibliography{references.bib}

\appendix

\section{Effects of probability cuts}
\label{appendix_P05}
In Section~\ref{sec:cont_eff}, we showed that a probability cut of $P_{\text{Ia}}>0.5$ can reduce contamination in the DES-SN sample by a factor of 4--5, depending on the \snn\ classifier considered. However, the BEAMS/BBC framework is specifically designed to handle samples that include both SNe Ia and contaminants, with the BEAMS likelihood calculated using $P_{\text{Ia}}$. Here we test the impact of combining BEAMS/BBC with probability-based cuts on cosmology.

\begin{figure}
    \includegraphics[width=\linewidth]{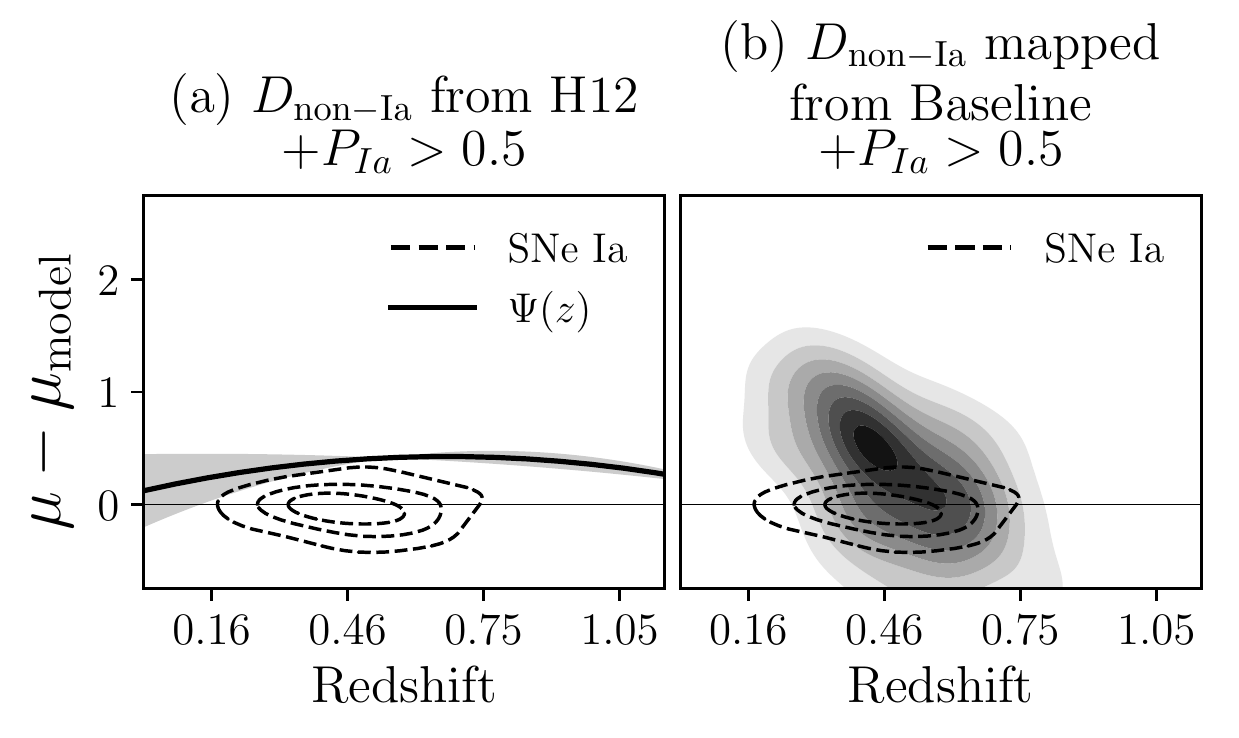}
\caption{Same as Fig.~\ref{fig:cc_maps} but when applying $P_{\mathrm{Ia}}$-based cuts.}
    \label{fig:cc_maps_pcut}
\end{figure}

\subsection{Core collapse SN likelihood and BBC configurations}

\begin{table}
\caption{BBC options tested with a SALT2 and  $P_{\mathrm{Ia}}$>0.5 selection. }
\label{table:BBC_opts_pcut}
\begin{tabular}{ccc}
\hline
BBC configuration & Classifier & Modelling \\
  &   & of $D_{\mathrm{CC}}$ \\

\hline
\texttt{SNN(Base)   D$_{\mathrm{CC}}$(Base) PIa>.5} &  \texttt{SNN(Base)} & Baseline \\
\texttt{SNN(J17) D$_{\mathrm{CC}}$(Base) PIa>.5} &  \texttt{SNN(J17)} &   Baseline \\
\texttt{SNN(H20)   D$_{\mathrm{CC}}$(Base) PIa>.5} &  \texttt{SNN(H20)} &   Baseline \\
\texttt{SNN(Base)   D$_{\mathrm{CC}}$(H12)  PIa>.5} &  \texttt{SNN(Base)} &   Fit (\citetalias{Hlozek_2012}) \\
\texttt{SNN(J17) D$_{\mathrm{CC}}$(H12)  PIa>.5} &  \texttt{SNN(J17)} &   Fit (\citetalias{Hlozek_2012}) \\
\texttt{SNN(H20)   D$_{\mathrm{CC}}$(H12)  PIa>.5} & \texttt{SNN(H20)} &   Fit (\citetalias{Hlozek_2012}) \\
\hline
\end{tabular}
\end{table}

\begin{figure*}
    \includegraphics[width=\linewidth]{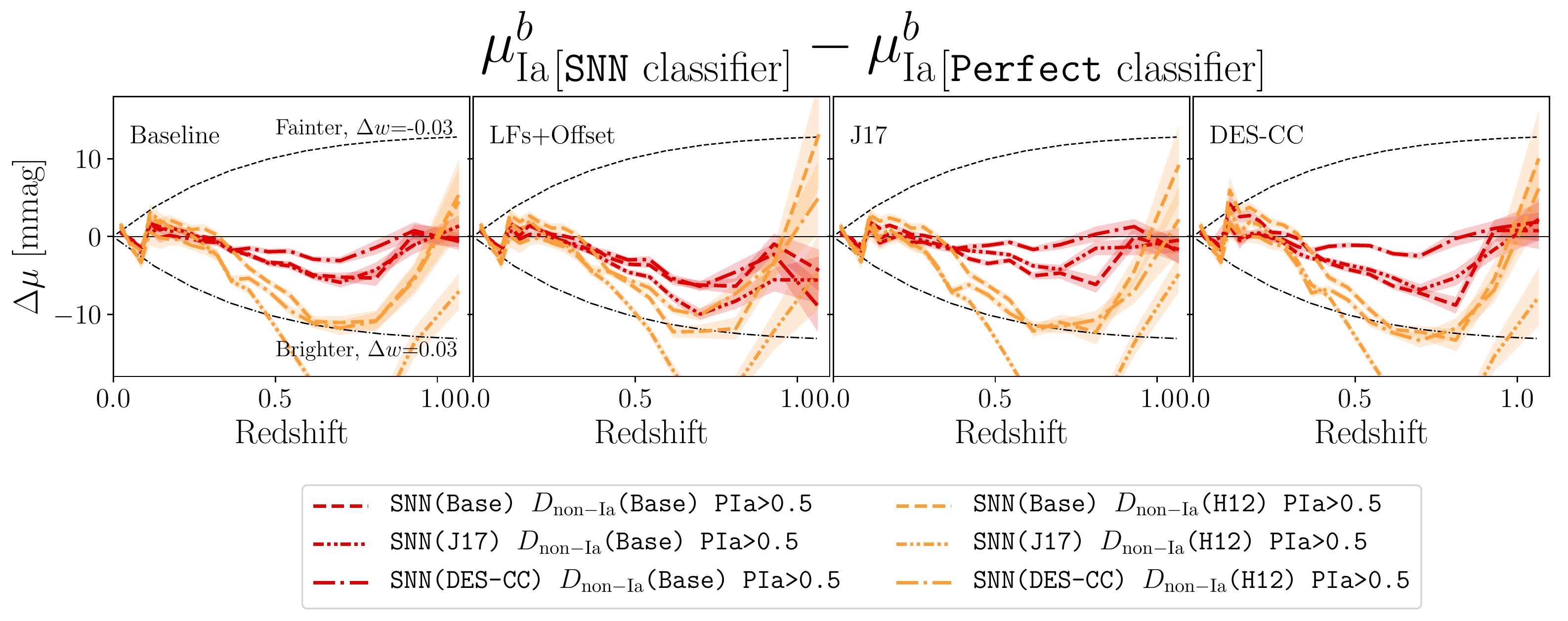}
    \caption{Same as Fig.~\ref{fig:compare_mudif}, but applying a $P_{\mathrm{Ia}}>0.5$ selection cut.}
    \label{fig:cont_eff_likelihood_Pcut}
\end{figure*}
\begin{figure*}
    \includegraphics[width=\linewidth]{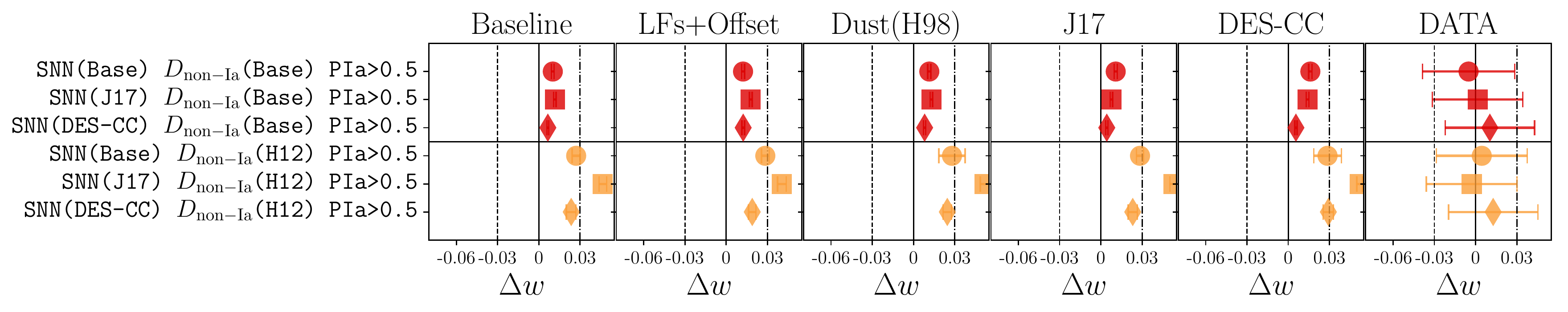}
\caption{Same as Fig.~\ref{fig:wbias_SNN} but for BBC configurations in Table~\ref{table:BBC_opts_pcut}.}
    \label{fig:wbias_SNN_pcut}
\end{figure*}

We combine the $P_{\mathrm{Ia}}>0.5$ selection and several different configurations of BBC, summarised in Table~\ref{table:BBC_opts_pcut}.
Applying a probability cut removes all SNe with $P_{\text{Ia}}<0.5$ from the main sample and from the simulations used to estimate bias corrections (which only include SNe Ia) and in the core-collapse SN simulation used to map the core-collapse SN likelihood ($\mathcal{L}_\text{CC}$). Probability cuts therefore have a complex impact on the analysis.

Fig.~\ref{fig:cc_maps_pcut} shows the effect of a probability selection on the two core-collapse SN likelihood models tested. Comparing Fig.~\ref{fig:cc_maps_pcut}a and Fig.~\ref{fig:cc_maps}a, the best fit $\Psi(z)$ (and relative $\sigma_{\text{CC, int}}(z)$) measured when a $P_{\mathrm{Ia}}>0.5$ selection is applied is significantly different from the best-fitting $\Psi(z)$ estimate without such a selection. This is expected because the $P_{\mathrm{Ia}}$ cut only selects the brightest contaminants and significantly reshapes the distribution of contamination on the Hubble diagram. This is particularly evident at high redshift, where contamination sharply drops and $\Psi(z)$ is an extrapolation.

Similar differences are seen when comparing the core-collapse SN maps derived from the Baseline simulation \textit{before} applying the $P_{\mathrm{Ia}}$ cut (Fig.~\ref{fig:cc_maps}b) and \textit{after} (Fig.~\ref{fig:cc_maps_pcut}b).
After $P_{\mathrm{Ia}}$ cuts, the distribution of core-collapse SN contamination is shifted to slightly higher redshifts ($0.3>z>0.7$), skewed towards the SN Ia likelihood (centred on $\mu_{\text{model}}$) and sharply reduced at high redshift.

\subsection{Conclusions}
Introducing a probability-based selection makes the modelling of $D_{\text{CC}}$ more complex. This can lead to significant biases when using the \citetalias{Hlozek_2012} approach, where the core-collapse SN likelihood is fitted from the data and it is assumed to be fully described by a second-order polynomial. After probability cuts, this assumption is not adequate because the contamination likelihood at high redshift is essentially an extrapolation of the fitted polynomial and it does not reflect the drop in contamination seen in simulations. 

In Fig.~\ref{fig:cont_eff_likelihood_Pcut}, we show that biases on fitted $\mu_{\mathrm{Ia}}^{\BinInd}$ when implementing \snn\, \texttt{$D_{\mathrm{non-Ia}}$(Base)} and $P_{\mathrm{Ia}>0.5}$ cut are still $<10$\,mag. However, when applying the \citetalias{Hlozek_2012} approach, the contamination likelihood is not robustly modelled and many high-redshift, faint SNe Ia are assigned a higher likelihood of being contaminants and excluded from the cosmological fit. This biases $\mu_{\mathrm{Ia}}^{\BinInd}$ towards negative values and propagates to the estimate of cosmological parameters. In Fig.~\ref{fig:wbias_SNN_pcut}, we show the $w$-biases for the different BBC configurations tested and we find biases larger than 0.04 for the majority of the configurations where the \citetalias{Hlozek_2012} approach is used. We note that the main driver of the bias in this case is not the presence of contaminants in the sample but the loss of SNe Ia in the cosmological fit.

Finally, when the contaminants likelihood is modelled from the Baseline simulation, the recovered biases are equal to or lower than 2--3 per cent and generally consistent with the biases found when a $P_{\mathrm{Ia}}$ cut is not applied. 

In summary, our tests show that applying a probability-based selection perhaps counter-intuitively provides equal or higher biases on cosmological parameters. The more accurate the classifier, the lower the residual contamination in the sample and the more uncertain the modelling of contamination in BEAMS. For these reasons, a probability-based selection is not recommended and we do not implement it in our main analysis.

\section{Rejecting SNe without a valid bias correction}
\label{sec:appendix_novalidbias}

\begin{figure}
\centering
\includegraphics[width=0.85\linewidth]{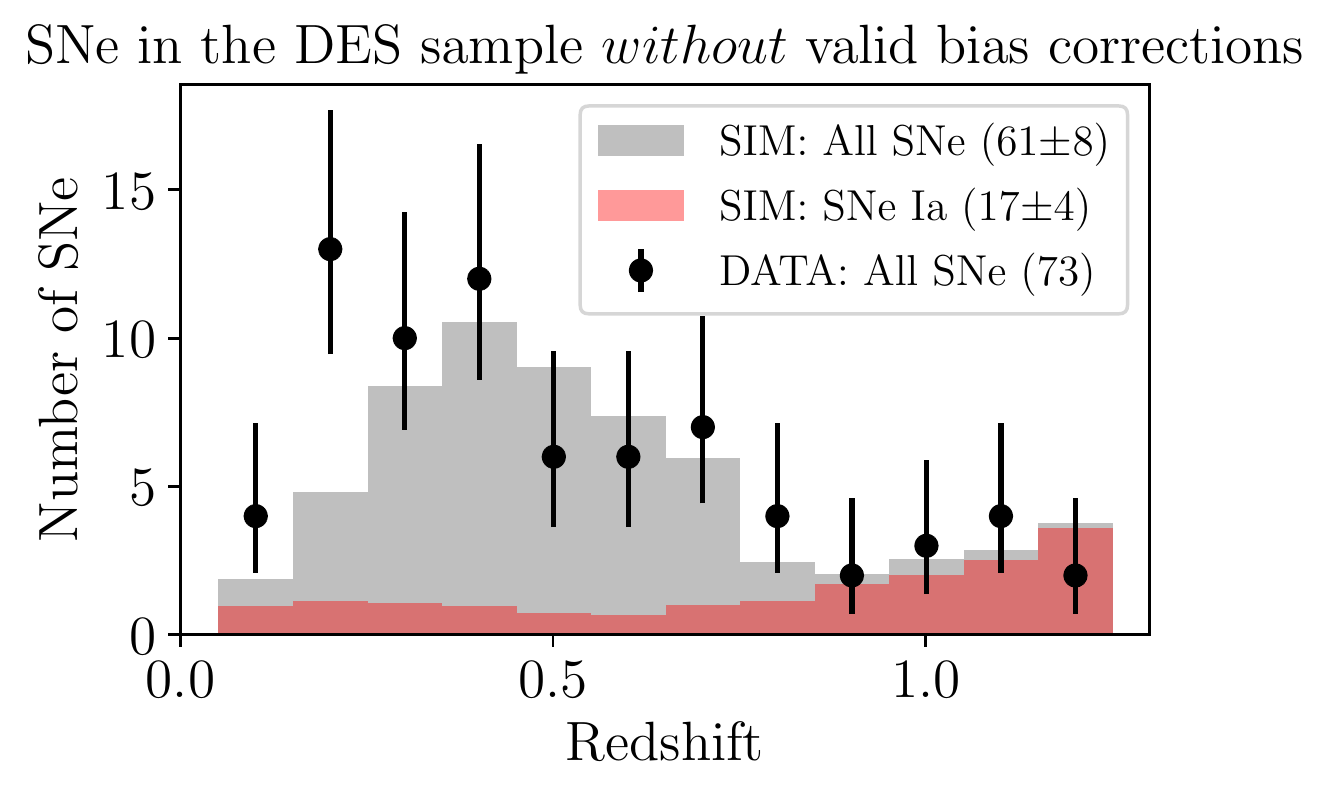}
\caption{Redshift distribution of SN events for which BBC does not provide valid bias corrections. We compare the sample of such events in the DES data (open histogram) with the sample of such events in our DES-like simulations (filled grey histogram). In the simulations, only a third of the SNe without valid bias corrections are SNe Ia (red filled histogram).}
    \label{fig:novalidBC}
\end{figure}

\begin{figure}
\centering
\includegraphics[width=0.8\linewidth,left]{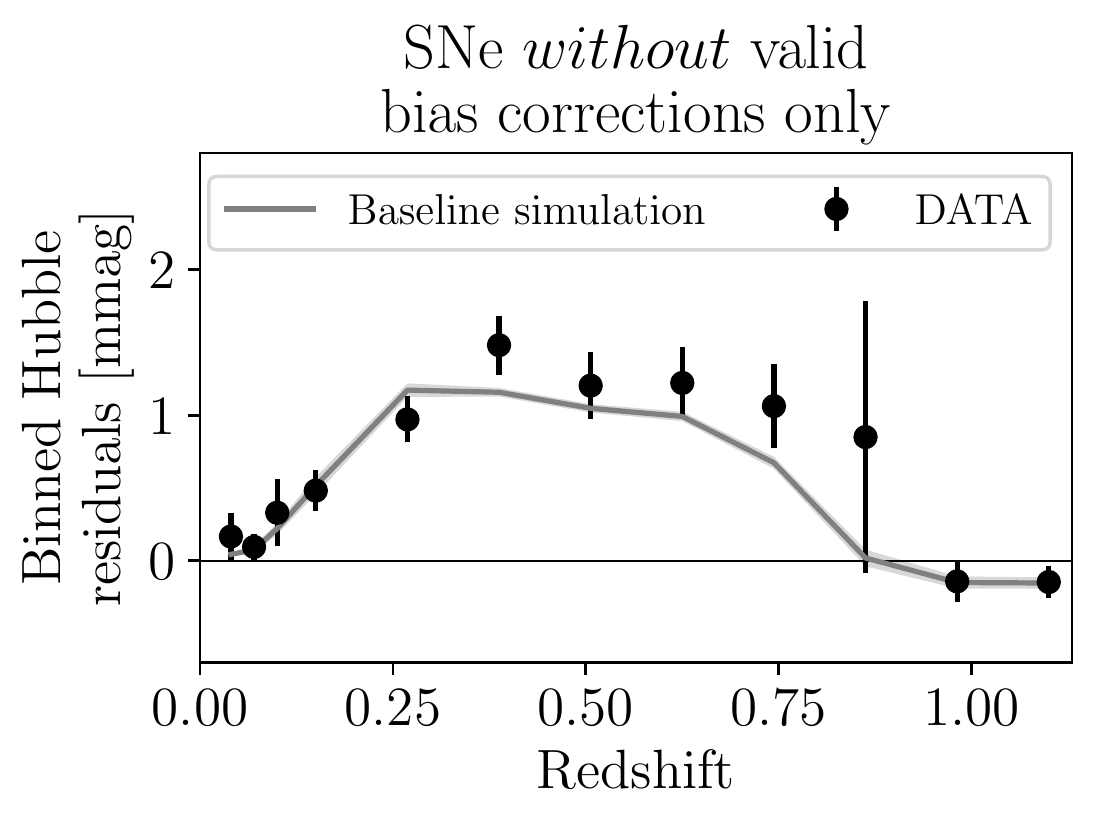}\\
\caption{Difference in the observed and simulated binned Hubble diagram estimated when using our reference BBC configuration but including SNe without valid bias corrections. Simulations are generated using the Baseline approach. Uncertainties are estimated as the r.m.s. spread measured over the 50 realizations of the Baseline simulation.}
    \label{fig:compare_mudif_chi2SNN_2nd}
\end{figure}

With BBC it is not always possible to estimate valid bias corrections for every SN, particularly those in regions of parameter space where few SNe are simulated (see Section~\ref{sec:biascor}). These SNe are excluded from a cosmological analysis and this reduces contamination in the sample. Here we test how our results change if the requirement of a valid bias correction is relaxed, and if SNe without a valid bias correction are retained in the sample but with $\mu_\mathrm{bias}$ set to zero. Since setting $\Delta \mu_{\mathrm{bias}}=0$ is clearly an incorrect approach in a cosmology analysis, we do not present updated results; rather, we compare the impact for data vs. simulation to ensure that this effect is properly modelled.

The requirement of a valid bias correction significantly affects both the low-$z$ and DES-SN samples (Table~\ref{table:cuts}), but here we focus on the DES-SN sample and effects at higher redshifts. In Fig.~\ref{fig:novalidBC}, we present the redshift distribution of observed and simulated DES SNe that pass the SALT2 selection, but do not have a valid bias correction. These distributions are generally consistent, but potential discrepancies are observed in the two lowest redshift bins. This suggests that the data include more atypical SNe than are modelled in the simulations.

Fig.~\ref{fig:compare_mudif_chi2SNN_2nd} shows the observed and simulated binned Hubble residuals estimated when considering only SNe without valid bias corrections. 
The average Hubble residuals of this sub-population of uncorrected events is low (less the 2 mmag) and consistent between observations and simulations. This test confirms that we can model the selection effects introduced by BBC, and it further validates the results obtained using our simulations. 
\section{AUTHOR AFFILIATIONS}
\label{aff}
$^{1}$ School of Physics and Astronomy, University of Southampton,  Southampton, SO17 1BJ, UK\\
$^{2}$ Institute of Cosmology and Gravitation, University of Portsmouth, Portsmouth, PO1 3FX, UK\\
$^{3}$ Department of Physics, Duke University Durham, NC 27708, USA\\
$^{4}$ Centre for Astrophysics \& Supercomputing, Swinburne University of Technology, Victoria 3122, Australia\\
$^{5}$ Universit\'e Clermont Auvergne, CNRS/IN2P3, LPC, F-63000 Clermont-Ferrand, France\\
$^{6}$ The Research School of Astronomy and Astrophysics, Australian National University, ACT 2601, Australia\\
$^{7}$ Centro de Investigaciones Energ\'eticas, Medioambientales y Tecnol\'ogicas (CIEMAT), Madrid, Spain\\
$^{8}$ African Institute for Mathematical Sciences, 6 Melrose Road, Muizenberg, 7945, South Africa\\
$^{9}$ Department of Maths and Applied Maths, University of Cape Town, Cape Town, South Africa\\
$^{10}$ South African Astronomical Observatory, Observatory, Cape Town, 7925, South Africa\\
$^{11}$ Center for Astrophysics $\vert$ Harvard \& Smithsonian, 60 Garden Street, Cambridge, MA 02138, USA\\
$^{12}$ NASA Einstein Fellow\\
$^{13}$ INAF, Astrophysical Observatory of Turin, I-10025 Pino Torinese, Italy\\
$^{14}$ School of Mathematics and Physics, University of Queensland,  Brisbane, QLD 4072, Australia\\
$^{15}$ Institut d'Estudis Espacials de Catalunya (IEEC), 08034 Barcelona, Spain\\
$^{16}$ Institute of Space Sciences (ICE, CSIC),  Campus UAB, Carrer de Can Magrans, s/n,  08193 Barcelona, Spain\\
$^{17}$ Department of Astrophysics, American Museum of Natural History, New York, NY, USA\\
$^{18}$ Department of Astronomy and Astrophysics, University of Chicago, Chicago, IL 60637, USA\\
$^{19}$ Kavli Institute for Cosmological Physics, University of Chicago, Chicago, IL 60637, USA\\
$^{20}$ Argonne National Laboratory, 9700 South Cass Avenue, Lemont, IL 60439, USA\\
$^{21}$ Sydney Institute for Astronomy, School of Physics, A28, The University of Sydney, NSW 2006, Australia\\
$^{22}$ Centre for Gravitational Astrophysics, College of Science, The Australian National University, ACT 2601, Australia\\
$^{23}$ Department of Physics and Astronomy, University of Pennsylvania, Philadelphia, PA 19104, USA\\
$^{24}$ Laborat\'orio Interinstitucional de e-Astronomia - LIneA, Rua Gal. Jos\'e Cristino 77, Rio de Janeiro, RJ - 20921-400, Brazil\\
$^{25}$ Fermi National Accelerator Laboratory, P. O. Box 500, Batavia, IL 60510, USA\\
$^{26}$ CNRS, UMR 7095, Institut d'Astrophysique de Paris, F-75014, Paris, France\\
$^{27}$ Sorbonne Universit\'es, UPMC Univ Paris 06, UMR 7095, Institut d'Astrophysique de Paris, F-75014, Paris, France\\
$^{28}$ Department of Physics \& Astronomy, University College London, Gower Street, London, WC1E 6BT, UK\\
$^{29}$ Kavli Institute for Particle Astrophysics \& Cosmology, P. O. Box 2450, Stanford University, Stanford, CA 94305, USA\\
$^{30}$ SLAC National Accelerator Laboratory, Menlo Park, CA 94025, USA\\
$^{31}$ Institut de F\'{\i}sica d'Altes Energies (IFAE), The Barcelona Institute of Science and Technology, Campus UAB, 08193 Bellaterra (Barcelona) Spain\\
$^{32}$ Astronomy Unit, Department of Physics, University of Trieste, via Tiepolo 11, I-34131 Trieste, Italy\\
$^{33}$ INAF-Osservatorio Astronomico di Trieste, via G. B. Tiepolo 11, I-34143 Trieste, Italy\\
$^{34}$ Institute for Fundamental Physics of the Universe, Via Beirut 2, 34014 Trieste, Italy\\
$^{35}$ Observat\'orio Nacional, Rua Gal. Jos\'e Cristino 77, Rio de Janeiro, RJ - 20921-400, Brazil\\
$^{36}$ Department of Physics, University of Michigan, Ann Arbor, MI 48109, USA\\
$^{37}$ Hamburger Sternwarte, Universit\"{a}t Hamburg, Gojenbergsweg 112, 21029 Hamburg, Germany\\
$^{38}$ Department of Physics, IIT Hyderabad, Kandi, Telangana 502285, India\\
$^{39}$ Santa Cruz Institute for Particle Physics, Santa Cruz, CA 95064, USA\\
$^{40}$ Institute of Theoretical Astrophysics, University of Oslo. P.O. Box 1029 Blindern, NO-0315 Oslo, Norway\\
$^{41}$ Instituto de Fisica Teorica UAM/CSIC, Universidad Autonoma de Madrid, 28049 Madrid, Spain\\
$^{42}$ Department of Astronomy, University of Michigan, Ann Arbor, MI 48109, USA\\
$^{43}$ Faculty of Physics, Ludwig-Maximilians-Universit\"at, Scheinerstr. 1, 81679 Munich, Germany\\
$^{44}$ Center for Cosmology and Astro-Particle Physics, The Ohio State University, Columbus, OH 43210, USA\\
$^{45}$ Department of Physics, The Ohio State University, Columbus, OH 43210, USA\\
$^{46}$ Australian Astronomical Optics, Macquarie University, North Ryde, NSW 2113, Australia\\
$^{47}$ Lowell Observatory, 1400 Mars Hill Rd, Flagstaff, AZ 86001, USA\\
$^{48}$ Observatories of the Carnegie Institution for Science, 813 Santa Barbara St., Pasadena, CA 91101, USA\\
$^{49}$ Department of Astrophysical Sciences, Princeton University, Princeton, NJ 08544, USA\\
$^{50}$ Departamento de F\'isica Matem\'atica, Instituto de F\'isica, Universidade de S\~ao Paulo, CP 66318, S\~ao Paulo, SP, 05314-970, Brazil\\
$^{51}$ George P. and Cynthia Woods Mitchell Institute for Fundamental Physics and Astronomy, and Department of Physics and Astronomy, Texas A\&M University, College Station, TX 77843,  USA\\
$^{52}$ Instituci\'o Catalana de Recerca i Estudis Avan\c{c}ats, E-08010 Barcelona, Spain\\
$^{53}$ Physics Department, 2320 Chamberlin Hall, University of Wisconsin-Madison, 1150 University Avenue Madison, WI  53706-1390\\
$^{54}$ Department of Astronomy, University of California, Berkeley,  501 Campbell Hall, Berkeley, CA 94720, USA\\
$^{55}$ Center for Astrophysical Surveys, National Center for Supercomputing Applications, 1205 West Clark St., Urbana, IL 61801, USA\\
$^{56}$ Institute of Astronomy, University of Cambridge, Madingley Road, Cambridge CB3 0HA, UK\\
$^{57}$ Department of Astrophysical Sciences, Princeton University, Peyton Hall, Princeton, NJ 08544, USA\\
$^{58}$ Computer Science and Mathematics Division, Oak Ridge National Laboratory, Oak Ridge, TN 37831\\
$^{59}$ Department of Physics, Stanford University, 382 Via Pueblo Mall, Stanford, CA 94305, USA\\
$^{60}$ Max Planck Institute for Extraterrestrial Physics, Giessenbachstrasse, 85748 Garching, Germany\\
$^{61}$ Universit\"ats-Sternwarte, Fakult\"at f\"ur Physik, Ludwig-Maximilians Universit\"at M\"unchen, Scheinerstr. 1, 81679 M\"unchen, Germany\\
$^{62}$ Department of Physics and Astronomy, Pevensey Building, University of Sussex, Brighton, BN1 9QH, UK\\

\bsp	
\label{lastpage}
\end{document}